\documentclass[preprint]{revtex4-2}
\nonstopmode
\usepackage{lmodern}
\usepackage{amsmath}
\usepackage{amsbsy}
\usepackage{pstricks,graphicx}
\RequirePackage{mathbbol}
\RequirePackage{amssymb}
\RequirePackage{mathtools}

\newcommand{\ket}[1]{\ensuremath{|#1\rangle}}
\newcommand{\braket}[2]{\ensuremath{\langle #1|#2\rangle}} 
\newcommand{\dirint}[3]{\ensuremath{\langle #1|#2|#3\rangle}}

\newcommand{\bs}{\boldsymbol}

\newcommand{\ul}{\underline}

\newcommand{\half}{{\textstyle\frac{1}{2}}}

\def\vcenterpdf#1{\mathrlap{\parbox{0pt}{\includegraphics[]{#1}}}\hphantom{\includegraphics[]{#1}}}

\begin{document}

\title{Review of the foundations of time-dependent density-functional theory (TDDFT)}
\author{J. Schirmer\\
Theoretische Chemie, Physikalisch-Chemisches Institut, Universit\"{a}t Heidelberg\\
Im Neuenheimer Feld 229, D-69120 Heidelberg, Germany}

\setcounter{tocdepth}{0}
\thispagestyle{empty}
\date{November 25, 2024}
\begin{abstract}

Time-dependent density-functional theory (TDDFT) is deemed to be a formally rigorous way of dealing with
the time-evolution of a many-electron system at the level of electron densities rather than the underlying wavefunctions, which in turn provides an, in principle, exact density-based approach to the treatment of electron excitations in atoms and molecules. This claim has not remained unchallenged, and a detailed account of the relevant criticism is given in this paper.  
In view of our analysis one has to face the conclusion that there is currently no valid justification for TDDFT, and expectations of finding a remedy here are hardly justified.
\end{abstract}
\maketitle

\section{Introduction}
Time-dependent density-functional theory (TDDFT) is a prominent approach
to the treatment of electronic excitations in many-electron systems. 
For comprehensive presentations and references, the reader is referred to a selection of more recent review 
articles~\cite{bur05:062206,ell08:91,cas09:3,gro12:53,rug15:203202,mai16:220901,hui21:15} and 
books~\cite{Engel:2011,Ullrich:2012}. A list of various TDDFT computer codes can be found in App.~O of the latter 
textbook. The computational performance of TDDFT variants has been examined 
in several benchmark studies~\cite{sil08:104103,jac09:2420,car10:370}.

From a pragmatic point of view, the TDDFT computational scheme can be seen as a time-dependent extension
of the Kohn-Sham (KS) equations of ground-state (static) DFT~\cite{hoh64:864,koh65:1133}. In fact, TDDFT was originally introduced in this way~\cite{zan80:1561}. However, a much bigger ambition is at stake here. In their famous 1984 paper~\cite{run84:997}, E.~Runge and E.~K.~U.~Gross (RG) have attempted to lay a rigorous foundation for the theory, thereby establishing the claim that, like DFT, TDDFT represents, in principle, an exact formulation for the time-evolution of an interacting many-electron system at the level of density functions.

The enormity of that claim, still widely considered valid today, should provoke profound scepticism. After all, in the time-evolution of quantum systems the phases of the wave functions are crucial, as indicated by the presence of the
imaginary factor $i$ in the time-dependent Schr\"{o}dinger equation. The positive definite real density functions, by contrast, are devoid of any phases, excluding even an innocent $(-1)$. Of course, in the time-dependent KS equations 
orbital-level phases come in through the backdoor, but the essential potential-functionals depend exclusively on the 
phase-free density functions. In view of this situation, it should be the natural reaction to ask where the error lies. 
On the other hand, any attempt to defend the status of TDDFT as an, in principle, exact theory should be based on rigorous mathematical and physical arguments, rectifying and transcending the original RG paper, which has been found defective. Regrettably, an actual follow-up founding paper is still outstanding. 

In 2007, a critical analysis was presented by Andreas Dreuw and the present author~\cite{sch07:022513} of the RG foundation of TDDFT and, moreover, of what was then considered a valid update of the RG concept, in which even a stationarity principle was no longer needed. This elicited a Comment by Maitra, van Leeuwen, and Burke~\cite{mai08:056501} and a subsequent Reply~\cite{sch08:056502},
further affirming and clarifying the original criticism. A certain conclusion was reached with a two-page  
paper~\cite{sch12:012514} in 2012, which, however, received little attention so far. 
Altogether, the main points of our criticism have been set out, albeit in a somewhat scattered and partly abridged form. Clearly, a comprehensive, consistent, and, moreover, pedagogical account of our survey would be desirable - and this paper 
attempts to fill that gap. It is based on a presentation given in May last year at the Joint Physical Chemistry Seminar at the Heidelberg University.

In the ensuing Sec.~II we consider the physics of electron excitations in many-electron systems, review the RPA (random-phase approximation) method and its benefit as a computational tool, and then take a closer look at the RPA-type linear-resonse (LR) TDDFT equations used in practice. In Sec.~III, the DFT essentials are 
briefly recapitulated, discussing, in particular, an utmost simplified (radical) Kohn-Sham scheme, which proves to be a useful analytical tool for DFT as well as TDDFT.  After these preliminaries, the actual criticism of the present TDDFT foundations is set out in Sec.~IV. A brief summary and concluding remarks are given in the final Sec.~V.

A methodological remark in advance may be appropriate. The intention of this paper is not to disprove the existence of TDDFT - indeed an impossible task, but rather identify and discuss unexpected gaps that invalidate its present foundation. Obviously, one does not need to strive for mathematical rigour to this end. 
By contrast, utmost rigour is required if one attempts to justify the validity of TDDFT - the burden of proof lies with the proponent of such an extraordinary claim.

\section{Electronic excitations and the RPA paradigm}

Before dealing with the foundation of TDDFT and, in particular, the justification of the 
time-dependent Kohn-Sham (td KS) equations, it is advisable to take a look at the 
computational procedure used in practice. Formally, the standard TDDFT scheme, referred to 
as linear response (LR) TDDFT is equivalent to the RPA (random-phase approximation) pseudo-eigenvalue
problem. This suggests to briefly recapitulate the RPA mathematics and its performance in
the application to electron excitations in molecules.

\subsection{Excited state physics}

In the one-particle picture, a singly excited state is obtained by promoting 
an electron from an occupied orbital, say $k$, to an unoccupied (or virtual) orbital, say $a$:
\begin{equation}
\ket{\Phi_{ak}}= c_a^\dagger c_k \ket{\Phi_0} 
\end{equation}
Here, $\ket{\Phi_0}$ is the 'unperturbed' Hartree-Fock (HF) ground state and 
$c_a^\dagger, c_k$ are the familiar Fermion field operators for the spin-orbitals $a, k$.
Obviously, $\ket{\Phi_{ak}}$ is an eigenstate,
\begin{equation}
\hat H_0 \ket{\Phi_{ak}} = E^{(0)}_{ak}\, \ket{\Phi_{ak}}
\end{equation}
of the HF (or zeroth-order) hamiltonian
\begin{equation}
\hat H_0 = \sum \epsilon_p c_p^\dagger c_p
\end{equation}
where $\epsilon_p$ denote the HF orbital energies;
$E^{(0)}_{ak}$ is the zeroth-order contribution to the total energy of the excited state.

The full hamiltonian can be partioned according to 
\begin{equation}
\hat H = \hat H_0 + \hat H_I
\end{equation}
into the zeroth-order part $\hat H_0$ and an interaction part, $\hat H_I  = \hat H - \hat H_0$, which
constitutes the starting point for perturbation-theoretical (PT) procedures.

More expedient than the total energy, $E_{ak}$, is the excitation energy, 
\begin{equation}
\Delta E_{ak} = E_{ak} - E_0
\end{equation}
where $E_0$ denotes the (exact) ground-state energy.
$\Delta E_{ak}$ can be evaluated through first order from the zeroth-order wave functions according to
\begin{equation}
\label{eq:dE1}
\Delta E_{ak}(1) = \dirint{\Phi_{ak}}{\hat H}{\Phi_{ak}} - \dirint{\Phi_0}{\hat H}{\Phi_0}
= \epsilon_a - \epsilon_k  - V_{akak} + V_{akka}
\end{equation}
The zeroth-order contribution is simply given by the difference of the HF orbital energies, which, however,
refer to the HF ground state. Accordingly, $\epsilon_a$ incorporates the Coulomb and exchange interaction with the 
orbital $k$, being vacant in the excited state. This is corrected by the first-order contribution, where the
Coulomb integral $V_{akak}$ accounts for the electron repulsion between the charge distributions of the $a$ and $k$ orbitals. The exchange integral, $V_{akka}$, leads to an energy-splitting of the singlet and triplet final states, being 
$2 V_{akka}$ (for $a, k$ with same spin) at the first-order level.

The exact state $\ket{\Psi_{ak}}$ associated with $\ket{\Phi_{ak}}$ can be computed via the standard procedure
referred to as configuration interaction (CI) in quantum chemistry.  
Here, the excited states are written as linear combinations 
\begin{equation}
\ket{\Psi_{n}} =\sum_{J} x_{Jn}\,\ket{\Phi_{J}}
\end{equation}
of HF configurations 
\begin{equation}
\ket{\Phi_{J}} = \hat C_J \ket{\Phi_0}
\end{equation}
where $\hat C_J$ denote excitation operators of the expansion manifold
\begin{equation}
\{\hat C_{J}\} = \{1; c_b^\dagger c_j; c_b^\dagger c_c^\dagger c_i c_j, b<c,i<j; \dots\}
\end{equation}
comprising single, double, triple, ... excitations. Note that the CI configurations also include the HF 
ground state, $\ket{\Phi_0}$.

The representation of the hamiltonian, or rather the shifted hamiltonian, $\hat H - E_0$,
in terms of the CI expansion manifold yields the CI secular matrix $\bs H$ with the elements
\begin{equation}
H_{IJ}=
\dirint{\Phi_{I}}{\hat{H}}{\Phi_{J}} - E_{0}\delta_{IJ}
\end{equation}
and the solution of the Schr\"{o}dinger equation is equivalent with solving the 
CI eigenvalue problem,
\begin{equation}
\bs{H}\bs{X}=\bs{X}\,\bs{\Omega},\quad \bs{X}^{\dagger}\bs{X}=\bs{1}
\end{equation}
Here, $\bs X$ is the matrix of eigenvectors, and $\bs \Omega$ is the diagonal
matrix of the eigenvalues, representing the excitation energies
\begin{equation}
\Omega_{n}=E_{n}-E_{0} 
\end{equation} 
Spectral intensities are related to the  
transition moments for a suitable operator $\hat D$, e.g., a component of the dipole operator,
\begin{equation}
T_{n}=\dirint{\Psi_{n}}{\hat{D}}{\Psi_{0}} = \sum_J X_{Jn}^* \dirint{\Phi_{J}}{\hat{D}}{\Psi_{0}}
\end{equation}

\begin{figure}
\begin{equation*}
\vcenterpdf{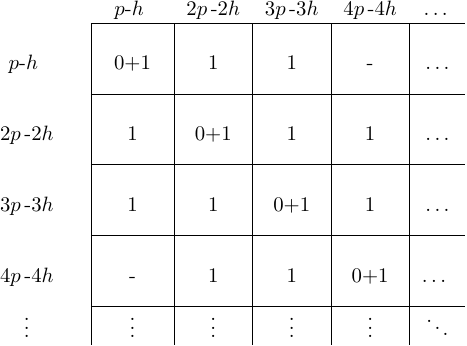}
\end{equation*}
\caption{Block structure of the CI secular matrix $\bs H$ (disregarding coupling to the HF ground state). 
The entries 0 and 1 refer to matrix elements involving $\hat H_0$ and $\hat H_I$, respectively.}
\label{fig:cicor}
\end{figure}

Fig.~\ref{fig:cicor} shows the block structure of the CI secular matrix and the PT order of the matrix elements in
the diagonal (0,1) and non-vanishing non-diagonal blocks (1). The coupling matrix elements $H_{IJ}$ vanish if the 
configurations $I$ and $J$ differ by more than two excitation classes.
For example, a single excitation such as $\ket{\Phi_{ak}}$ can have non-zero coupling matrix elements with $p$-$h$,
$2p$-$2h$, and $3p$-$3h$ configurations but not with configurations of higher excitation classes. 
This, in turn, implies that the second-order contribution to the excitation energy takes on the form
\begin{equation}
\label{eq:dE2}
\Delta E^{(2)}_{ak} = U^{(2)}_{ak}(p\text{-}h) +  U^{(2)}_{ak}(2p\text{-}2h) +
U^{(2)}_{ak}(3p\text{-}3h) - E^{(2)}_0
\end{equation}
Here, the first term, $U^{(2)}_{ak}(p\text{-}h)$, is due to admixtures of other single excitations,  possibly effecting a better adapted virtual orbital $a$ (as well as the occupied orbital $k$), or even reflecting a genuine mixed $p$-$h$ composition of the final state.  
The second term, $U^{(2)}_{ak}(2p\text{-}2h)$, is of particular physical significance, as it accounts for the 
\textbf{relaxation} and \textbf{polarization} effects accompanying the $k \rightarrow a$ excitation. The remaining electrons ``relax" upon the removal of an electron from the orbital $k$; and the electron in the virtual orbital $a$ ``polarizes" the charge distribution of the ionic core. Together, these two response effects lead to a substantial lowering of the 
first-order excitation energy. 
For an explicit PT analysis 
of the relaxation and polarization energies the reader is referred to Ref.~\cite{sch95:105}.

The third term, 
\begin{equation}
U^{(2)}_{ak}(3p\text{-}3h) = 
 - \sum_{\substack{b<c\neq a\\i<j\neq k}} \frac{|V_{bc[ij]}|^2}{\epsilon_b \!+\!\epsilon_c\!-\!\epsilon_i\!-\!\epsilon_j}
\end{equation}
where $V_{pq[rs]}$ denotes anti-symmetrized Coulomb integrals,
accounts for the correlation effect in the excited state. It is of the same form as the familiar second-order 
PT expression, $E_0^{(2)}$, for the ground state energy, though with restrictions for the orbital summation indices.
Obviously, the second-order correlation energy is larger in the ground state than in the excited state, so that together 
the last two terms, giving rise to the explicit expression
\begin{align}
\nonumber
 U^{(2)}_{ak}(3p\text{-}3h) &- E^{(2)}_0  = \\
\label{eq:umine} 
&\sum_{b,i<j} \frac{|V_{ab[ij]}|^2}{\epsilon_a\!+\!\epsilon_b\!-\!\epsilon_i\!-\!\epsilon_j} 
+ \sum_{j,b<c} \frac{|V_{bc[kj]}|^2}{\epsilon_b\! +\! \epsilon_c\!-\!\epsilon_k\!-\!\epsilon_j} 
-\sum_{j,b} \frac{|V_{ab[kj]}|^2}{\epsilon_a\!+\!\epsilon_b\!-\!\epsilon_k\!-\!\epsilon_j}  
\end{align}
are positive, effecting an increase of the excitation energy. Note that the negative third contribution on the right-hand side, called $-\,C$ for short, compensates for the fact that a positive counterpart, $+\,C$, is 
comprised both in the first and second contribution in Eq.~(\ref{eq:umine}).
 
An adequate description of the first- and second-order physics just outlined
should be considered a standard requirement that any reliable computational approach to electron excitations 
must fulfill. In the case of the CI method, this means that the expansion manifold must not be truncated before 
the $3p$-$3h$ (triple) excitations, which entails rather large configuration spaces (compactness problem). 
However, the inclusion of the triples causes another problem inherent to truncated CI expansion (as opposed to full CI) schemes, that is, the well-known size-consistency problem (see e.g. Ref.~\cite{Schirm:2018}). 

In this section we have considered the case of single excitations. It should be noted, however, that double and higher excitations can be addressed in a similar fashion.

\subsection{Random-phase approximation (RPA)}

In theoretical terms, the RPA~\cite{mcl64:844,dun67:1735,row68:153} is a highly interesting concept, whereas its 
computational benefit in the application to electron excitations in molecules is rather modest.  
For a detailed presentation and references the reader may consult Chapter 15 of Ref.~\cite{Schirm:2018}.
There are various independent derivations of the RPA, one of which is of particular interest in the present context,
namely the derivation based on the time-dependent Hartree-Fock (TDHF) approach. 
For a presentation of the TDHF route to the RPA equations, the reader is referred to the textbook by 
Ring and Schuck~\cite{Ring:1980}.

The RPA secular equations constitute a pseudo-eigenvalue problem 
of the form
\begin{equation}
\left( \begin{array}{cc}
                      \bs{A}   &  \bs{B}\\
                      \bs{B}^* & \bs{A}^*
         \end{array} 
         \right)
                     \left( \begin{array}{c}
                                  \ul{x}_m\\
                                  \ul{y}_m
                                  \end{array}
                            \right) = 
                         \omega_m \left( \begin{array}{c}
                                  \ul{x}_m\\
                                   - \ul{y}_m
                                  \end{array}
                            \right)
\end{equation} 
Here $\bs A$ and $\bs B$ are sub-blocks of a hermitian matrix $\bs{\mathcal M}$ defined 
with respect to a configuration manifold of $p$-$h$ excitations and $h$-$p$ `de-excitations'.
The matrix elements are given by
\begin{align}
\nonumber
A_{ak,bl} =& (\epsilon_a - \epsilon_k) \delta_{ab} \delta_{kl} -V_{albk} + V_{alkb}\\
B_{ak,lb} =& -V_{ablk} + V_{abkl}
\end{align}
The $p$-$h$/$p$-$h$ block $\bs A$ can readily be identified with the CI-S (singles) secular matrix.
The role of the $\bs B$ block and the unfamiliar coupling between $p$-$h$ and $h$-$p$ configurations 
will be discussed below.

There is a manifold of excitation (pseudo-) eigenvalues,  
\begin{equation}
\omega_m = E_m - E_0
\end{equation}
representing physical excitation energies
and a manifold of redundant de-excitation solutions, $\omega_m^- = - \omega_m^*$.
The transition moments are obtained according to
\begin{equation}
T_m = \sum_{a,k} \left (x_{ak,m}^* d_{ak} + y_{ka,m}^* d_{ka}\right)
\end{equation}
from the $p$-$h$ and $h$-$p$ eigenvector components of the excitation solutions and transition operator matrix 
elements, $d_{pq}$.

To gain some insight into the essence of the RPA pseudo-eigenvalue problem it is instructive
to consider the $2\times 2$ RPA-type eigenvalue problem
\begin{equation*}
\left( \begin{array}{cc}
                      a   &  b\\
                      b &  a
         \end{array} 
         \right)
                     \left( \begin{array}{c}
                                  x\\
                                  y
                                  \end{array}
                            \right) = 
                         \omega \left( \begin{array}{c}
                                  x\\
                                   -y
                                  \end{array}
                            \right)
\end{equation*} 
where $a$ and $b$ are real numbers.
The two eigenvalues are given by
\begin{equation*}
\omega_{\pm} = \pm \sqrt{a^2-b^2} = \pm |a| \sqrt{(1 - b^2/a^2)}
\end{equation*} 
The corresponding RPA eigenvectors can easily be determined as well. 
In the case $|b| > |a|$, complex eigenvalues result. 
Supposing $|b| < |a|$ the square root can be expanded in powers of $b^2/a^2$, 
yielding the PT expansion
\begin{equation*}
\omega_+  =  a - \frac{b^2}{2a} + \dots 
\end{equation*}
for $\omega^+$, here supposing $a>0$. Note that the PT denominator is given by the sum of the diagonal elements.

As in the preceding Section II.1, the RPA results can be further analyzed 
by inspecting their PT expansions through second-order.
The RPA excitation energy for the $k \rightarrow a$ single excitation  reads 
\begin{equation}
\Delta E^{\scriptscriptstyle RPA}_{ak} = \epsilon_a  -\epsilon_k -  V_{akak} + V_{akka}
+ U^{(2)}_{ak}(p\text{-}h) + U^{(2)}_{ak}(h\text{-}p) + O(3)
\end{equation}
The comparison with Eqs.~(\ref{eq:dE1},\ref{eq:dE2}) shows that the RPA excitation energy is consistent through 
first order (which also applies to the transition moment), whereas in second order only a subset of the full 
expression is recovered. While the $U^{(2)}_{ak}(p\text{-}h)$ term is accounted for,  
the physically more important $U^{(2)}_{ak}(2p\text{-}2h)$ term is absent, 
which means that the relaxation and polarization effects are disregarded.
The other second-order RPA term, $U^{(2)}_{ak}(h\text{-}p)$, arises due to the unusual coupling of
the considered $a$-$k$ excitation with $h$-$p$ de-excitation configurations.
The explicit expression, given by
\begin{equation}
U^{(2)}_{ak}(h\text{-}p) =  -\sum_{b,j} \frac{V^2_{ab[kj]}}{\epsilon_a \! + \! \epsilon_{b} \!
- \!\epsilon_k \! - \! \epsilon_{j}}
\end{equation}
is identical with the third term on the right-hand side of Eq.~(\ref{eq:umine}).
This means that the RPA scheme accounts for the negative correction term, referred to as $-\,C$, in the overall positive correlation contribution to the excitation energy. 
While the lowering of the excitation energy thereby entailed may to a certain extent compensate for the 
lack of the (negative) relaxation and polarization shift, the resulting improvement of the computational results
is not based on physical grounds.

As the PT analysis shows, the RPA method can hardly be seen as a satisfactory approach to electron excitations.
The treatment is restricted to single excitations, and here the resulting excitation energies and transition moments are consistent only through first order of PT. Due to the absence of double excitations, the essential effects of 
relaxation and polarization are not taken into account. And, finally, there is an improper consideration of 
ground and final-state correlation effecting a reduction rather than an increase of the excitation energies.

\subsection{LR-TDDFT computational scheme}

Analogous to the connection of the RPA with TDHF, 
the LR-TDDFT computational scheme is based on the time-dependent KS equations.
Supposing the almost exclusively used adiabatic approximation (time-dependence via 
time-dependent density functions),     
LR-TDDFT equations can be derived in an obvious modification of the TDHF/RPA procedure, 
as, e.g., discussed in Ref.~\cite{Ring:1980}. 
Of course, independent derivations within the TDDFT framework have been reported  
as well~\cite{cas95:155}.  

The LR-TDDFT computational scheme (in the adiabatic approximation) is formally equivalent to the 
RPA pseudo-eigenvalue method discussed above. The excitation energies, 
\begin{equation}
\omega_m = E_m - E_0
\end{equation}
and transition moments, 
\begin{equation}
T_m = \sum_{b,l} \left (x_{bl,m}^* d_{bl} + y_{lb,m}^* d_{lb}\right)
\end{equation}
are given by the eigenvalues and eigenvector components, respectively,
of the RPA-type secular equations
\begin{equation}
\label{eq:lr-tddft}
\left( \begin{array}{cc}
                      \tilde{\bs{A}}   &  \tilde{\bs{B}}\\
                      \tilde{\bs{B}}^* & \tilde{\bs{A}}^*
         \end{array} 
         \right)
                     \left( \begin{array}{c}
                                  \ul{x}_m\\
                                  \ul{y}_m
                                  \end{array}
                            \right) = 
                         \omega_m \left( \begin{array}{c}
                                  \ul{x}_m\\
                                   - \ul{y}_m
                                  \end{array}
                            \right)
\end{equation} 
Here, the matrix elements of the $\tilde{\bs{A}}$ and $\tilde{\bs{B}}$ blocks read
\begin{align}
\nonumber
\tilde A_{ak,bl} =& (\epsilon_a^{\scriptscriptstyle KS}- \epsilon_k^{ \scriptscriptstyle KS}) \delta_{ab} \delta_{kl} +  f^{xc}_{alkb}+ V_{alkb} \\
\label{eq:lr-tddftx}
\tilde B_{ak,lb} =&  f^{xc}_{abkl} + V_{abkl}
\end{align}
These expressions 
differ from their RPA counterparts by using the KS orbital energies rather than HF energies
and by replacing the Coulomb integrals $-V_{albk}$ in $\bs A$ by the matrix elements 
\begin{equation}
f^{xc}_{alkb} = \dirint{\phi_a\phi_{l}}{f^{xc}[n_0]}{\phi_{k}\phi_{b}}
\end{equation}
and, similarly, 
$-V_{ablk}$ in $\bs B$ by $f^{xc}_{abkl}$. 
Here 
\begin{equation}
f^{xc}[n](\bs r, \bs r') = \frac{\delta  v_{xc}[n](\bs r)}{\delta n (\bs r')}
\end{equation}
denotes the so-called xc kernel, that is, the functional derivative of the xc potential 
functional, $v_{xc}[n](\bs r)$,
and $n_0(\bs r)$ is the initial density of the system (before the onset of the td perturbation).
Explicit expressions for adiabatic xc functionals have been given, for example, by Bauernschmitt and Ahlrichs~\cite{bau96:454}.

It should be noted that the LR-TDDFT expressions~(\ref{eq:lr-tddftx}) suppose the so-called adiabatic approximation,
in which the xc functionals depend on time only via the time-dependence of the densities, $n(t)$. This allows one to resort to potential functionals associated with the ground-state energy functionals, $E_{xc}[n]$. 
For local functionals, such as LDA (local density approximation) or GGA (generalized gradient approximation) functionals, the
functional derivatives become local quantities as well, 
\begin{equation}
f^{xc}[n_0](\bs r, \bs r') \sim f(\bs r)\delta(\bs r -\bs r')
\end{equation}
so that the matrix elements $f^{xc}_{pqrs}$ are reduced to  
one-particle integrals.

Again, the LR-TDDFT results can be inspected via RPA-type PT for the 
excitation energy through second order, yielding 
\begin{equation}
\Delta E^{\scriptscriptstyle TDDFT}_{ak} = \epsilon^{\scriptscriptstyle KS}_a -\epsilon^{\scriptscriptstyle KS}_k   +
f^{xc}_{akka} + V_{akka}
+ \tilde U^{(2)}_{ak}(p\text{-}h) + \tilde U^{(2)}_{ak}(h\text{-}p) + O(3)
\end{equation}
The comparison with Eq.~(\ref{eq:dE1}) shows that already the simple first-order physics 
is compromised, as there is no longer the crucial Coulomb repulsion term, $-V_{akak}$, 
accounting for the Coulomb repulsion of the electrons in the orbitals $a$ and $k$.
Whether $f^{xc}_{akka}$ can be seen as 
an appropriate replacement is questionable. At least, for local functionals the approximate $1/R$ dependance of the 
original Coulomb integral on the distance $R$ between localized orbital $a$ and $k$ cannot be recovered. 
Obviously, this is the cause for the problems arising in the LR-TDDFT treatment of Rydberg-type excitations~\cite{cas98:4439,toz98:10180} and charge transfer (CT) 
excitations~\cite{toz99:859,dre03:2943,dre04:4007,sob03:73}.  
The problem of the RPA in dealing with the response effects accompanying the $a$-$k$ excitation, that is, 
relaxation and polarization, seems to affect also the LR-TDDFT results. There are no double excitations, and 
it can hardly be expected that their default can otherwise be compensated for.  
Finally, one may wonder whether the $\tilde U^{(2)}_{ak}(h\text{-}p)$ term may be any better than the RPA analogue
in dealing with the ground- and final state correlation energies.

In conclusion, the LR-TDDFT can be seen as a modification of the RPA computational scheme, providing 
physically reasonable excitation energies (and transition moments) for singly excited states. However, one can probably argue whether
LR-TDDFT affords an significant improvement over the RPA description and whether eventual improvements are based on 
physical grounds. Of course, the more profound question is: Where do the td KS equations actually come from in the first place?

\section{Review of DFT}

Conceptually, TDDFT draws largely upon the well-established DFT approach to the 
ground-state energy and density. This warrants a brief recapitulation of the DFT essentials. In particular,
we will discuss an elementary form of the Kohn-Sham (KS) scheme, which can also be used as 
an analytical tool in the case of TDDFT.  
For thorough and detailed presentations of DFT the reader is referred to  monographs~\cite{Parr:1989,Engel:2011} as well as review articles (see e.g. Nagy~\cite{nag98:1} and references therein).

\subsection{Hohenberg-Kohn theorems}
 
Specifically, we consider an N-electron system with the hamiltonian given by
\begin{equation}
\hat{H} =\hat{T} + \hat{V} + \hat{U}
\end{equation} 
where 
$\hat{T}$ is the kinetic energy, 
$\hat{V}$ is the electron repulsion, and $\hat U$ denotes a local one-particle potential, e.g., the electron-nuclei 
interaction potential,
\begin{equation}
\hat{U} = \sum_{i=1}^N u(\bs{r}_i),\,\,\,u(\bs{r}) = 
- e^2 \sum_{a=1}^K \frac{Z_a}{|\bs{R}_a -\bs{r}|}
\end{equation}
Let 
\begin{equation}
\Psi = \Psi(\bs{r}_1 s_1, \dots,\bs{r}_{N} s_{N})
\end{equation}
denote an antisymmetrized and normalized wave function, where $\bs r_i$ and $s_i$ are the spatial coordinates and
spin variable, respectively, of the $i$-th electron.  
The associated energy can be obtained as the
expectation value, 
\begin{equation}
E[\Psi] = \dirint{\Psi}{\hat{H}}{\Psi}
\end{equation} 
The ground-state energy, $E_0$, is given by the lowest eigenvalue of the (static) 
Schr\"{o}dinger equation,
\begin{equation}
\hat{H}\Psi_0 = E_0\Psi_0
\end{equation}
which, in turn, is fully equivalent with a 
variational principle according to
\begin{equation}
\delta E[\Psi_0] = 0, \quad \braket{\Psi_0}{\Psi_0} = 1
\end{equation}

The DFT approach is based on the much simpler (one-particle) density functions, $n(\bs r)$, 
deriving from the original $N$-electron wave functions according to 
\begin{equation}
n(\bs{r}) = N \sum_{s_1,\dots,s_N}\int \dots \int |\Psi(\bs{r}s_1, \bs{r}_2s_2,\dots)|^2 d\bs{r}_2\dots d\bs{r}_N
\end{equation}
Obviously, the density functions are positive definite, $n(\bs{r}) \geq 0$, and normalized according to
\begin{equation*}
\int n(\bs{r}) d\bs{r} = N
\end{equation*}

Can the energy directly be deduced from the density function $n(\bs r)$ rather than the underlying wave function $\Psi$? 
Indeed, it is the basic DFT tenet that an energy functional can be established, 
\begin{equation}
E = E[n]
\end{equation}
mapping a given density function to the corresponding energy. However, 'establish' here means a proof of existence
of such a functional, whereas a definite construction procedure is not available and one ultimately has to resort to appropriate guesses.
Let us inspect the individual constituents of the hamiltonian.
Obviously, the energy contribution associated with the local one-particle potential $\hat U$, also referred to as the external potential can directly be computed from the density function: 
\begin{equation}
U[n] = \int u(\bs{r}) n(\bs{r}) d\bs{r}
\end{equation}
This does not apply to the kinetic energy part $T[n]$, as here a density-based evaluation would require 
the one-particle density matrix. In the electron repulsion energy $V[n]$ one may distinguish the classical Coulomb energy 
\begin{equation}
J[n] = \half \, e^2 \int \int 
\frac{n(\bs{r})n(\bs{r}')}{|\bs{r} - \bs{r}'|} d\bs{r} d\bs{r}'
\end{equation}
of a charge distribution $e\, n(\bs r)$ and a (non-classical) remainder 
\begin{equation}
\label{eq:xcfun}
E_{xc}[n] = V[n] - J[n]
\end{equation}
accounting for the exchange effect (anti-symmetrization of the $N$-electron wave function) and the correlation effect
(deviation of the wave function from the Slater product form).
Like the kinetic energy, this non-classical remainder $E_{xc}[n]$, referred to as exchange-correlation energy, cannot directly be derived at the one-particle density level.

However, according to the first theorem of Hohenberg and Kohn (HKI)~\cite{hoh64:864}, one can, at least, establish the existence of the needed functionals. The argument goes as follows. There is an obvious mapping of (local) external potentials, $v(\bs r)$, to one-particle densities
as indicated in the following mapping scheme:
\begin{equation}
\label{eq:v2n}
 v(\bs{r}) \longrightarrow \hat{H}[v] \longrightarrow \Psi[v] \longrightarrow n[v](\bs{r})
\end{equation}
Here, 
\begin{equation}
\hat{H}[v] = \hat T + \hat V + \sum_{i=1}^N v(\bs r_i)
\end{equation}
is the $N$-electron hamiltonian associated with the external potential $v(\bs r)$; $\Psi[v]$ is the ground-state
wave function of $\hat{H}[v]$, and $n[v](\bs{r})$ is the one-particle density function associated with $\Psi[v]$.
Now, the crucial point is that this mapping is essentially one-to-one: If two potentials are mapped to the same density function, 
they must be identical up to a constant. As a consequence, the mapping~(\ref{eq:v2n}) can be inverted,  
\begin{equation}
\label{eq:n2v}
n(\bs{r}) \longrightarrow v[n](\bs{r}) + c
\end{equation}
Mathematical strictness demands to mention that here the domain of eligible densities may be restricted to so-called 
$v$-\textit{representable} densities.
Continuing the inverse mapping with the original mapping sequence~(\ref{eq:v2n}) 
according to
\begin{equation}  
n(\bs{r}) \longrightarrow v[n](\bs{r}) + c \longrightarrow \hat{H}[n] + C \longrightarrow \Psi[n]
\end{equation}
establishes a mapping between the densities $n(\bs r)$ and the wave functions $\Psi[n]$, being the ground-state wave function of the hamiltonian $\hat H[n] = \hat H[v[n]]$ with the external potential $v(\bs r) = v[n](\bs r)$.
This allows one, in particular, to establish the 
universal HK functional according to
\begin{equation}
\label{eq:hkfun1}
F[n] = \dirint{\Psi[n]}{\hat{T} + \hat{V}}{\Psi[n]}
\end{equation}
Herewith the desired 
energy functional for the considered system becomes
\begin{equation}
\label{eq:hkfun2}
E[n] =  F[n] + \int n(\bs{r}) u(\bs{r}) d\bs{r}
\end{equation}
This functional is subject to a variational principle,
\begin{equation}
\label{eq:vapri}
E_0 = E[n_0] \leq E[n]
\end{equation}
where $E_0$ and $n_0(\bs r)$ are the ground-state energy and density function, respectively. This is the essence of the 
second Hohenberg-Kohn theorem (HKII). 

The variational principle can be translated into Euler-Lagrange equations briefly considered below. This offers
a shortcut to $E_0$ and $n_0$ bypassing the Schr\"{o}dinger equation. Again, it must be emphasized 
that the HKI theorem establishes the existence of $E[n]$ without providing a means for its construction.
 
It is worth noting that the energy functionals can also be 
established in a more direct way referred to as Levy's constrained search (see Parr and Yang~\cite{Parr:1989}).
According to the definition
\begin{equation}
E[n]  =  \min_{\Psi \rightarrow n} \, \dirint{\Psi}{\hat{H}}{\Psi} 
  =  \min_{\Psi \rightarrow n} \, \dirint{\Psi}{\hat{T} +\hat{V}}{\Psi} + \int u(\bs{r}) n(\bs{r}) d\bs{r}
\end{equation}
one samples the energy expectation values with respect to all (normalized) $N$-electron wave functions
compatible with a given density function $n(\bs r)$ and takes the minimal value limit.
The universal HK functional, for example, is given by
\begin{equation*}
F[n] = \min_{\Psi \rightarrow n} \, \dirint{\Psi}{\hat{T} +\hat{V}}{\Psi}
\end{equation*}
Note that according to the LCS definition the variational principle,
\begin{equation}
E_0 = E[n_0] \leq E[n]
\end{equation}
is manifest. The densities are required to be $N$-\textit{representable}, which is supposed to be less restrictive than the $v$-representability required in the HK mapping approach.

Supposing the exact or an approximate energy functional is available, 
the variational principle,
\begin{equation}
E_0 = E[n_0] \leq E[n] \;\; \text{for}\;\; \int n(\bs r) d\bs r = N
\end{equation}
allows one to determine 
the ground-state density, $n_0(\bs r)$, and energy, $E_0$, 
via the Euler-Lagrange equation
\begin{equation}
\label{eq:eula}
\delta E[n] - \mu \int \delta n(\bs{r}) d\bs{r} = 0
\end{equation}
where $\mu$ is a Lagrange parameter.
The variation $\delta E[n]$ can be expressed in terms of the functional derivative,
\begin{equation}
\delta E[n] = \int \frac{\delta E[n]}{\delta n(\bs{r})} \delta n(\bs{r}) d\bs{r}
\end{equation}
Consequently, Eq.~(\ref{eq:eula}) can be written in the form
\begin{equation}
\label{eq:eleq}
\frac{\delta E[n]}{\delta n(\bs{r})} = \mu
\end{equation}

According to Eqs.~(\ref{eq:xcfun},\ref{eq:hkfun1},\ref{eq:hkfun2})
the energy functional comprises four parts,
\begin{equation}
E[n] = T[n] + E_{xc}[n] + J[n] + \int u(\bs{r})n(\bs{r})d\bs{r}
\end{equation}
which is reflected by a corresponding partitioning of the total functional derivative:
\begin{equation}
\frac{\delta E[n]}{\delta n(\bs{r})} =
\frac{\delta T[n]}{\delta n(\bs{r})} + \frac{\delta E_{xc}[n]}{\delta n(\bs{r})}
+ J[n](\bs r) + u(\bs{r})  \\
\end{equation}
Here,
\begin{equation}
J[n](\bs r) =  e^2 \int 
\frac{n(\bs{r}')}{|\bs{r} - \bs{r}'|} d\bs{r}'
\end{equation}
is the explicit functional derivative of the classical Coulomb energy $J[n]$.

As a computational scheme, Eq.~(\ref{eq:eleq}) is hardly of practical use. It represents a 
3-dimensional integral equation, which, like the Thomas-Fermi equation, may be tractable for atoms,
where the rotational symmetry can be exploited, but the application to molecules would be a serious challenge. 
Moreover, as density functions 
do not constitute a Hilbert space (as the difference of two density functions may no longer be a density function), one has to forgo basis set expansions and other mathematical conveniences.
This is where the Kohn-Sham concept comes into play.

\subsection{Kohn-Sham method}

The basic measure in the Kohn-Sham (KS) version~\cite{koh65:1133} of DFT is to 
introduce an orbital level 
beneath the densities. According to the mapping  
\begin{equation}
\Phi = |\psi_1 \psi_2 \dots \psi_N|\,\,\, \longrightarrow\,\, 
n (\bs{r}) = \sum_{i,s} |\psi_i(\bs{r},s)|^2 = 2 \sum_{k}^{N/2} |\phi_k(\bs r)|^2
\end{equation}
a given density can be obtained from the Slater determinant $|\psi_1 \psi_2 \dots \psi_N|$
of the spin-orbitals $\psi_{i} = \phi_k \chi_\gamma, k = 1, \dots, N/2; \gamma = \alpha, \beta$, or, directly, via the spatial orbitals $\phi_k, k = 1, \dots, N/2$.
By definition, the representation
\begin{equation} 
 n = n\{\psi_i\}
\end{equation}
herewith established, applies to 'non-interacting N-representable densities',
which, however, is hardly a restriction at all. It should be noted, though, that
the repesentation is not necessarily unique.

The recourse to the orbital level allows one to define an approximate kinetic energy expression 
according to 
\begin{equation}
T_{KS}[n\{\psi_i\}] \equiv \sum_{i,s}  \int \psi^*_{i}(\bs{r},s)(-\half \nabla^2) \psi_{i}(\bs{r},s)\, d\bs{r} = 2 \sum_k^{N/2} \int \phi^*_{k}(\bs{r} )(-\half \nabla^2) \phi_{k}(\bs{r})\, d\bs{r}
\end{equation}
and shift (or dispose of) the remainder, $T[n] - T_{KS}[n]$, in the 
exchange-correlation functional, now defined as
\begin{equation}
\label{eq:ksxcfun}
E_{xc}[n] = V[n] - J[n] + T[n] - T_{KS}[n]
\end{equation}
Accordingly, the original energy functional can be written as 
\begin{equation}
\label{eq:ksenfun}
E[n\{\psi_i\}]  =  T_{KS}[n\{\psi_i\}] + J[n] + E_{xc}[n] + \int u(\bs{r}) n (\bs{r})\, d\bs{r}
\end{equation}
and the task of finding the minimum $E[n_0]$ can now be performed at the orbital level,
that is, in Hilbert space. 
Like in the familiar derivation of the Hartree-Fock (HF)  or self-consistent field (SCF) equations, the variation with respect to the 
spin-orbitals $\psi_i$, or likewise, the spatial orbitals $\phi_k$, maintainig their orthonormality,
\begin{equation} 
\delta E[n\{\psi_i\}] = 0, \,\,\,\braket{\psi_i}{\psi_j} = \delta_{ij}
\end{equation}
gives rise to the set of Kohn-Sham equations, which in spin-free form read
\begin{equation}
\label{eq:kseq}
\{-\half \nabla^2 + u(\bs{r}) + J[n](\bs{r}) 
                   + v_{xc}[n](\bs{r})\}\, \phi_k(\bs{r}) = \epsilon_k \phi_k(\bs{r}),\,\, k=1,\dots,N/2
\end{equation}
Here, 
\begin{equation}
 v_{xc}[n](\bs{r}) = \frac{\delta E_{xc}[n]}{\delta n(\bs{r})}
\end{equation} 
is the exchange-correlation potential (more precisely, xc potential functional), obtained as the functional derivative of the xc energy functional.
Like the SCF equations, the KS equations~(\ref{eq:kseq}), together with the relation
\begin{equation}
n(\bs r) = 2 \sum_{k}^{N/2} |\phi_k(\bs r)|^2
\end{equation}
establish an iterative computational scheme, devised to converge to the ground-state density, $n_0(\bs r)$, 
marking the minimum of the energy functional. The ground-state energy is obtained as $E_0 = E[n_0]$ via Eq.~(\ref{eq:ksenfun}).
 
To summarize once again, the KS formulation must be seen as a mathematical artifice rather than a physically 
motivated construction. This means, in particular, that the KS orbitals and orbital energies have no
theoretically founded physical meaning (except perhaps for the highest occupied orbital~\cite{lev84:2745}); they may acquire a certain significance due to the 
similarity between the KS and SCF computational procedures, which extends to the respective orbitals.

\subsection{Radical Kohn-Sham scheme}

In the preceding section, we have characterized the KS method as a mathematical device for 
determining the minimum of the DFT energy functional at the orbital level, allowing one to operate in Hilbert space. 
Once this has been understood, one needs not to stop at the original KS variant with its representation of   
the density in terms of $N$ non-interacting particles. 
Any particle number is admitted, even $N=1$, that is, the 
representation of the density associated with a single (spinless) particle! The whole KS machinery can readily be 
transferred to such a representation referred to as radical Kohn-Sham (rKS) scheme in Ref.~\cite{sch07:022513}.
Under different designations, such KS modifications have previously been considered by 
Levy et al.~\cite{lev84:2745} and Holas and March~\cite{hol91:5521}.

Any given density $n(\bs{r})$ can trivially be assigned
to an orbital,
\begin{equation}
\label{eq:rks1}
\phi (\bs{r}) = \left (\frac{n (\bs{r})}{N}\right )^{1/2}
\end{equation}
and the density, in turn, is related to that orbital according to 
\begin{equation} 
\label{eq:rks2}
n(\bs{r}) = N \phi (\bs{r})^2
\end{equation}
This establishes an explicit one-to-one mapping of density functions and (real) orbitals. 

An obvious choice for the kinetic energy at the orbital level is 
\begin{equation}
\label{eq:rks3}
\widetilde T_{KS}[n\{\phi\}] =  N \int \phi(\bs{r})(-\half \nabla^2)\phi (\bs{r})\, d\bs{r}
\end{equation}  
Using the correspondingly modified xc functional
\begin{equation}
\widetilde{E}_{xc}[n] = V[n] - J[n] + T[n] - \widetilde T_{KS}[n] 
\end{equation}
the original energy functional can be written as
\begin{equation}
E[n]  =   \widetilde T_{KS}[n] + J[n] + \widetilde{E}_{xc}[n] + \int u(\bs{r}) n (\bs{r})\, d\bs{r}
\end{equation}

Via Eq.(\ref{eq:rks2}) and (\ref{eq:rks3}), 
the energy functional is defined at the orbital level, $E[n] = E[n\{\phi\}]$,
and the variational principle can be exploited with respect to 
$\phi$, 
\begin{equation*}
\delta E[n\{\phi\}] = 0, \quad \braket{\phi}{\phi} = 1
\end{equation*}
This results in the 
single KS equation
\begin{equation}
\label{eq:rkseq}
\{-\half \nabla^2 + u(\bs{r}) + J[n](\bs{r}) 
                   +  \widetilde{v}_{xc}[n](\bs{r})\}\, \phi(\bs{r}) = \epsilon \phi(\bs{r})
\end{equation}                   
where
\begin{equation*}
\widetilde{v}_{xc}[n](\bs{r}) = \frac{\delta \widetilde{E}_{xc}[n]}{\delta n(\bs{r})} 
\end{equation*}
is the modified xc potential functional.
As in the regular KS scheme, the single rKS equation~(\ref{eq:rkseq}) and Eq.~(\ref{eq:rks2})
establish a fixed-point iteration (FPI) scheme  
for the ground-state density $n_0(\bs r)$ and energy $E_0 = E[n_0]$. 

As the usual KS method, the rKS formulation is, in principle, exact. 
In the rKS version the essentially mathematical nature of the KS concept is manifest, which applies  
also to the significance of the KS orbitals: There is only a single `occupied' KS orbital here.   
In view of its large deviation from the SCF computational scheme, the rKS  cannot be recommended for practical use. 
However, it proves to be a useful tool for analytical and pedagogical purposes.  
In this capacity, its td extension will be instrumental to clarify certain issues of the TDDFT venture.

\section{Foundations of TDDFT}

\subsection{Original design by Runge and Gross}

The basic constituent of the TDDFT foundation by Runge and Gross~\cite{run84:997}
is a theorem (RG1) which establishes a time-dependent analogue to the HK1 theorem, namely,
a mapping of time-dependent densities to time-dependent potentials. 
This, in turn, allows one to define an action-integral functional (AIF). 
Together with a stationarity principle, to be seen as a reasonable postulate, a density-based equation-of-motion (EOM)
can be established governing the time-evolution of the $N$-electron system under consideration.

Let us inspect the RG1 theorem. For a given time-dependent density, $n(\bs r,t)$, in a time interval, say
$t_0\leq t \leq t_1$,
there is a time-dependent `external' potential,
\begin{equation}
\label{eq:rg1x}
n(\bs r,t) \longrightarrow v_{ext}[n](\bs{r},t) + \; c(t), \; t_0\leq t \leq t_1
\end{equation}
defined up to a purely time-dependent function, $c(t)$,
such that the corresponding $N$-electron time-dependent (td) Schr\"{o}dinger equation (SE)
\begin{equation}
\label{eq:rg1y}
i \frac{\partial}{\partial t} \Psi[n](t)  = 
\{ \hat{T} + \hat{V} + \hat{V}_{ext}[n](t)\} \Psi[n](t), \;\; \hat{V}_{ext}[n](t) = \sum_i v_{ext}[n](\bs{r}_i,t) + C(t)
\end{equation}
reproduces the original density: 
\begin{equation}
\label{eq:rg1z}
\Psi[n](t) \longrightarrow  n(\bs{r},t)
\end{equation}
Obviously, the mapping depends on the initial conditions at $t=t_0$, which, however, can be 
assumed to be appropriately specified, e.g., according to $\Psi[n](t_0) = \Psi[n_0]$, where $n_0(\bs r) = n(\bs r,t_0)$
and $\Psi[n_0]$ conforms to the HKI mapping. 
The external potential $\hat V_{ext}[n]$ is uniquely determined up to a purely time-dependent function, $C(t)$,
and consequently $\Psi[n](t)$ is only defined up to a phase factor, $e^{-i\alpha(t)}$.

Now, we again consider a specific $N$-electron system, where the 
hamiltonian,
\begin{equation}
\hat{H}(t) = \hat{T}+ \hat{V} + \hat{U}(t) 
 \end{equation} 
features also a time-dependent local (external) potential. It may be supposed that
the time-dependent contribution sets in at $t_0$, that is, the one-particle potential is of the form
\begin{equation}
u(\bs r,t) = u(\bs r) + \theta(t-t_0) w(\bs r,t) 
\end{equation} 
Starting from a given wave-function $\Psi_0$, e.g., the ground-state wavefunction of the static part of the hamiltonian,
the solution of the td SE provides the time-dependent wave function, $\Psi_0(t)$, and thereof the corresponding density trajectory, $n_0(\bs r,t)$. 
This is the standard procedure of quantum theory. TDDFT, by contrast, claims that $n_0(\bs r,t)$ can be obtained without solving the td SE. How is that supposed to work?

There are two steps in the RG foundation. The first is the 
definition of an \textbf{action-integral functional} (AIF) for the system under consideration based on the RG1 theorem:
\begin{equation}
\label{eq:aif}
A[n] = 
 \int_{t_0}^{t_1} dt\, 
\dirint{\Psi[n] (t)}{i\frac{\partial}{\partial t} -  \hat H(t)}{\Psi[n] (t)}
 \end{equation} 
Here, $\Psi[n] (t)$ is the wave function associated with the density function $n(t)$ according to the 
RG1 mapping equations~(\ref{eq:rg1x}-\ref{eq:rg1z}).
And, second, there is a \textbf{stationarity principle}
postulating that $A[n]$ is stationary at the sought density function $n_0(\bs r,t)$, 
\begin{equation}
\label{eq:aifsp} 
\delta A[n] = \int_{t_0}^{t_1} dt \int d \bs r \,
\frac{\delta A[n]}{\delta n (\bs r,t)}\,\delta n (\bs r,t) = 0 
\end{equation} 
This implies the equation
\begin{equation}
\frac{\delta A[n]}{\delta n (\bs r,t)} = 0
\end{equation} 
for the time-dependent functional derivative of $A[n]$, which
is supposed to be the desired EOM 
for the density trajectory $n_0(\bs r,t)$.

In their 1984 paper, RG do not elaborate on that time-dependent Euler-type equation, but move on to 
establish a 'practical scheme', that is, time-dependent KS equations.  
However, as a closer examination has revealed~\cite{sch12:012514}, there is a fatal problem with 
the AIF as defined in Eq.~(\ref{eq:aif}). 
According to the RG1 construction,  
$\Psi[n](t)$ fulfills the td SE~(\ref{eq:rg1y}),
\begin{equation}
i \frac{\partial}{\partial t} \Psi[n](t)  =
\{\hat{T} + \hat{V} + \hat{V}_{ext}[n](t)\} \Psi[n](t)
\end{equation}
Using this in Eq.~(\ref{eq:aif}),
the AIF simply becomes 
\begin{align}
\nonumber
A[n] =& 
 \int_{t_0}^{t_1} dt\, 
\dirint{\Psi[n](t)}{i\frac{\partial}{\partial t} -  \hat H (t)}{\Psi[n](t)} \\
\nonumber
=&  \int_{t_0}^{t_1} dt 
\dirint{\Psi[n](t)}{\hat{V}_{ext}[n](t) - \hat U(t)}{\Psi[n](t)} \\
= &  \int_{t_0}^{t_1} dt \int d\bs r 
\left\{v_{ext}[n](\bs r,t) - u(\bs r,t)\right \} n(\bs r,t)
\end{align}
This explicit form shows that the RG AIF is an essentially trivial construct. The
time derivative has disappeared, which means that the stationarity principle~(\ref{eq:aifsp})
cannot result in a time-dependent EOM. Moreover, there is no longer any direct reference to the kinetic energy, $\hat T$, and the
Coulomb repulsion, $\hat V$, of the considered system.    
The remaining content is what is already implied by the RG1 mapping theorem, namely,
\begin{equation}
\label{eq:rgsv}
v_{ext}[n_0](\bs r,t) = u(\bs r,t) + c(t)
\end{equation}
so that $A[n_0] = 0$, supposing here that the undefined time function $c(t)$ can be ignored. 
Obviously, 
the particular RG1 mapping result is not a means to determine the desired density $n_0(\bs r, t)$. 

Problems associated with the RG AIF were realized as early as 1996, as one encountered the  
so-called causality paradox, that is, the finding that the kernel of the xc functional 
is not consistent with causality~\cite{Nalewajski:1996,raj96:3916,lee98:1280,har99:5101}.
Regrettably, the cause of the failure was not ultimately analysed and communicated. Instead, several explanations
emerged. 
As suggested by van Leeuwen~\cite{lee01:1969}, the problem is due to the arbitrary td phase in $\Psi[n](t)$ 
that renders the AIF ill-defined. This stance was also adopted in our 2007 paper~\cite{sch07:022513} - though now being outshone by the finding in Ref.~\cite{sch12:012514}.
Vignale~\cite{vig08:062511} located the problem in the boundary conditions of the variation and 
proposed a corresponding remedy. The view that 
together with the Vignale correction the RG AIF is still viable can be found in the recent TDDFT literature~\cite{hui21:15}. One also comes across the view that ultimately there are no problems here at all~\cite{Engel:2011}. 

In fact, a clarification of the AIF issue was no longer seen a topmost priority. 
The general position adopted by the TDDFT architects was that the AIF concept is dispensable, 
since a rigorous foundation of the theory could be based entirely on RG1-type mapping theorems.
A rigorous foundation of a density-based theory for the time-evolution of quantum systems without
a stationarity principle? This calls for a closer examination.

\subsection{TDDFT based only on RG1-type mappings?}

Via the RG1 mapping a potential functional is established,
by which an external potential is assigned to a given (td v-representable) density, 
\begin{equation*}
n(\bs{r},t) \longrightarrow  v_{ext}[n](\bs{r},t)
\end{equation*}
so that the td SE for the interacting many-electron system reproduces the density $n(\bs{r},t)$ (in a given interval
$t_0 \leq t \leq t_1$, supposing appropriate initial conditions).
A corresponding mapping scheme can also be constructed for a system of non-interacting particles~\cite{lee99:3863}: 
\begin{equation}
n(\bs{r},t) \longrightarrow  v_{KS}[n](\bs{r},t)
\end{equation}
Here, $v_{KS}[n](\bs{r},t)$ is referred to as KS potential (functional). Using $v_{KS}[n](\bs{r},t)$ in the one-particle  td
Schr\"{o}dinger equations,
\begin{equation}
i \frac{\partial}{\partial t} \psi_k(\bs{r},t) = 
\{-\half \nabla^2 + v_{KS}[n](\bs{r},t)\} \psi_k(\bs{r},t), \; k=1,\dots,N/2
\end{equation}
reproduces the density according to
\begin{equation}
 n(\bs{r},t) = 2 \sum_k|\psi_k(\bs{r},t)|^2
\end{equation}
(Note that here the initial conditions for the orbitals can be chosen according to $\psi_k(\bs r,0) = \phi_k(\bs r)$,
where $\phi_k(\bs r)$ are the KS orbitals associated with the density $n(\bs r,0)$.)

So there are two RG1-type mapping schemes: one for interacting and the other for non-interacting particles. 
It is believed that a valid foundation of TDDFT can be obtained as the result of a particular interplay of
these two RG1 variants. 
Before taking a closer look at this in the following Sec.~IV.D,  
we shall consider a time-dependent extension of the rKS scheme
discussed in Sec.~III.C for static DFT. 
Here, the (non-interacting) mapping,
\begin{equation}
n(\bs{r},t) \longrightarrow  w[n](\bs{r},t)
\end{equation}
being based on a single orbital td equation,
can explicitly be constructed and analyzed, as will be discussed in the following.

\subsection{Time-dependent radical KS scheme}

Let $n (\bs{r},t)$ be a density in the time interval $t_0\leq t \leq t_1$. 
A time-dependent orbital $\psi(\bs r,t)$
associated with $n(\bs{r},t)$, that is, $n(\bs{r},t) = N |\psi(\bs r,t)|^2$,
is of the general form  
\begin{equation}
\label{eq:tdo} 
\psi[n](\bs{r},t) =\phi(\bs{r},t) e^{ik(\bs{r},t)},  \;\;  
\phi(\bs{r},t) = \left (\frac{n(\bs{r},t)}{N}\right )^{1/2}
\end{equation}
Here, the modulus of $\psi(\bs r,t)$ is directly established by $n (\bs{r},t)$,
but even the phase function, $k(\bs{r},t)$, is determined by the density up to a purely time-dependent function. 
The latter follows from the requirement that $\psi(\bs r,t)$ is the solution of 
a one-orbital td SE of the type 
\begin{equation}
i \frac{\partial}{\partial t} \psi(\bs{r},t) = \{-\half \nabla^2 + w(\bs{r},t)\} \psi(\bs{r},t)
\end{equation}
and therefore the continuity equation applies (see Ref.~\cite{sch07:022513}):
\begin{equation}
\label{eq:conteq}
\dot n(\bs r,t) + \nabla \cdot \bs{j}(\bs r,t) = 0
\end{equation}
Here, $\dot n$ denotes the time derivative of $n$, and $\bs j(\bs r,t)$ is the current density given by
\begin{equation}
\label{eq:conteqx}
\bs{j}(\bs r,t) = n(\bs r,t)\, \nabla k(\bs r,t)
\end{equation}
The first of the latter two equations 
allows one to determine (via vector field analysis) $\bs j(\bs r,t)$ in terms of $\dot{n}$. The second equation then 
leads to $\nabla k$, which, in turn, determines $k(\bs r,t)$ up to a constant at a given time, or, overall, up to a purely time-dependent function, $\alpha(t)$.
Altogether, this establishes a direct and essentially explicit mapping,  
\begin{equation}
\label{eq:dem}
n(\bs{r},t) \longleftrightarrow  \psi[n](\bs{r},t) = \left (\frac{n(\bs{r},t)}{N}\right )^{1/2}\,
                e^{ik[n](\bs{r},t)} \, e^{i \alpha (t)}
\end{equation}
between density trajectories, $n(\bs r,t)$, and time-dependent orbitals, $\psi[n](\bs{r},t)$.  
It should be noted that this also relates to the initial state, $\psi[n](\bs{r},t_0)$.

This mapping of densities and orbitals can readily be used to 
formulate the first stage of the  
RG1-type mapping,
\begin{equation}
n (\bs{r},t) \longleftrightarrow  w[n](\bs{r},t)
\end{equation}
that is, the mapping
between densities and td KS-type potentials, here for a single orbital td SE,
\begin{equation}
\label{eq:sotde}  
i \frac{\partial}{\partial t} \psi[n](\bs{r},t) = 
\{-\half \nabla^2 + w[n](\bs{r},t)\} \psi[n](\bs{r},t) 
\end{equation}
To this end, one can simply insert the expression given in Eq.~(\ref{eq:dem})
for $\psi[n](\bs{r},t)$. The real part can be solved for $w[n](\bs{r},t)$, 
\begin{equation}
\label{eq:wnt} 
w[n](\bs{r},t) = \frac{\nabla^2 \phi[n]}{2 \phi[n]} - \half (\nabla k[n])^2  - \dot{k}[n] - \dot \alpha(t)
\end{equation}
whereas the imaginary part reproduces the continuity equation~(\ref{eq:conteq}). 

Using the latter expression for $w[n]$ in
the single-orbital td SE~(\ref{eq:sotde}),
\begin{equation} 
i \frac{\partial}{\partial t} \psi(\bs{r},t) = 
\{-\half \nabla^2 +  w[n(t)](\bs{r},t)\} \psi(\bs{r},t)
\end{equation} 
reproduces the density according to 
\begin{equation}
n (\bs{r},t) = N |\psi(\bs{r},t)|^2
\end{equation}
where the time-dependent orbital, $\psi(\bs{r},t)$, is
consistent with the expression~(\ref{eq:tdo}).

The explicit construction of the RG1-type mapping for a one-orbital system
shows some interesting features.   
Foremost, we note that the potential functional, $w[n](\bs r,t)$, is \textbf{non-instantaneous}.
To construct the phase function $k[n](\bs r,t)$ at a given time $t$,
both the density, $n(t)$, and its time derivative, $\dot n(t)$, are required (at time $t$). 
According to Eq.~(\ref{eq:wnt}), the time derivative $\dot k[n]$ enters the expression 
for $w[n](\bs r,t)$. This means that $n(t)$, $\dot{n}(t)$, and $\ddot{n}(t)$ are needed
in the construction of the rKS potential $w[n](\bs r,t)$ at a given time $t$. 
This feature must be expected as well in the KS potential functional $v_{KS}[n(t)]$ 
arising in the context of a non-interacting many-particle system.

Another observation, pertaining both to the KS and rKS case, is that the KS potential functionals
are \textbf{trivial} insofar as they reproduce (via the orbital td SE) any given density,
including the desired density $n_0(\bs r,t)$ of the  
system under consideration:
\begin{equation}
v_{KS}[n_0(t)](\bs{r},t)  \,\,\, \longrightarrow \,\,n_0(\bs{r},t)
\end{equation}
Obviously, this is not a means to determine $n_0(\bs{r},t)$, since the latter density trajectory
(together with its first and second time derivatives) must be available in order to construct 
the needed potential $v_{KS}[n_0(t)](\bs{r},t)$.

\subsection{Time-dependent Kohn-Sham equations as a fixed-point iteration scheme}

Now we can come back to the issue addressed in Sec.~IV.B: How can, based on the RG1-type mapping theorems,
the td KS equations actually be established? 
Unfortunately, this issue has not been spelled out with utmost clarity in the TDDFT literature.  
A common presentation is as follows 
(see e.g. Refs.~\cite{mar04:427,ell08:91,Ullrich:2012,mai16:220901}):
All the `magic' is already in the KS potential functional (PF) established via the RG1-type  mapping for non-interacting systems. 
So one may simply rewrite it according to 
\begin{equation}
\label{eq:ofver}
v_{KS}[n](\bs{r},t) \equiv u(\bs{r},t) + J[n](\bs{r},t) + 
         v_{xc}[n](\bs{r},t)
\end{equation}
where $u(\bs{r},t)$ is the one-particle potential of the system under consideration and $J[n](\bs{r},t)$ is the Hartree
PF, and use it in the td orbital equations, 
\begin{equation}
i\frac{\partial}{\partial t} \psi_k(\bs{r},t) = 
\{-\half \nabla^2  + u(\bs{r},t) + J[n](\bs{r},t) +
 v_{xc}[n](\bs{r},t)\} \,\psi_k(\bs{r},t), \; k=1,\dots,N/2
\end{equation}
which then just look like the familiar td KS equations. 
However, as we have seen, the KS PF is a trivial rather than a 'magic' PF, that is, it reproduces any given
density, but is not a means to find the particular density $n_0(\bs r,t)$ of the given system.
How could that change by simply rewriting the original PF? Moreover, Eq.~(\ref{eq:ofver}) seems to 
establish a definition of the xc PF, $v_{xc}[n](\bs{r},t)$, which is a bit strange. The KS PF on the left-hand side is 
a universal entity, as are $v_{xc}[n](\bs{r},t)$ and $J[n](\bs{r},t)$ on the right-hand side, whereas 
$u(\bs{r},t)$ is the particular potential of the given system.

Clearly, this version cannot be the answer, and, indeed, there is a
more satisfactory substantiation, as outlined, for example, in Ref.~\cite{rug15:203202}. 
Based on the RG1-type mappings for both the interacting and the non-interacting systems,
a non-trivial xc PF can be defined according to
\begin{equation}
\label{eq:xcpfnt}
 v_{xc}[n](\bs{r},t)\equiv v_{KS}[n](\bs{r},t) - v_{ext}[n](\bs{r},t) 
- J[n](\bs{r},t)
\end{equation}
Now, td KS equations of the form
\begin{align}
\nonumber
i\frac{\partial}{\partial t} \psi_k(\bs{r},t) =& \{-\half \nabla^2  + u(\bs{r},t) + J[n](\bs{r},t) +
 v_{xc}[n](\bs{r},t)\} \,\psi_k(\bs{r},t), \; k=1,\dots,N/2 \\
\label{eq:tdksfp}
n(\bs r,t) =& 2 \sum_k^{N/2} |\psi_k(\bs{r},t)|^2
\end{align}
can simply be \textbf{postulated}. So what is the rationale here?
Actually, Eqs.~(\ref{eq:xcpfnt},\ref{eq:tdksfp}) establish a sort of a fixed-point
iteration (FPI) scheme for td densities ($t_0 \leq t \leq t_1$), where the sought density $n_0(\bs r,t)$ is the fixed point.

This can be seen as follows.
Using the definition~(\ref{eq:xcpfnt}) for the xc PF, the td KS equations can be written more explicitly as 
\begin{equation}
i\frac{\partial}{\partial t} \psi_k(\bs{r},t) = \{-\half \nabla^2  + u(\bs{r},t) + v_{KS}[n](\bs{r},t) - 
               v_{ext}[n](\bs{r},t)\} \,\psi_k(\bs{r},t),\; k= 1,\dots, N/2
\end{equation}
Recall that for $n_0(\bs r, t)$ the RG1 mapping yields
\begin{equation}
v_{ext}[n_0](\bs{r},t) = u(\bs{r},t) + c(t)
\end{equation}
This means that for $n(\bs r,t) \rightarrow n_0(\bs r,t)$ 
the external potential cancels the specific potential $u(\bs{r},t)$ so that 
the td KS equations take on the form
\begin{equation}
i\frac{\partial}{\partial t} \psi_k(\bs{r},t) = 
\{-\half \nabla^2 + v_{KS}[n_0](\bs{r},t)\} \,\psi_k(\bs{r},t)\;\longrightarrow \; n_0(\bs{r},t)
\end{equation}
where finally the KS PF $v_{KS}[n_0](\bs{r},t)$ reproduces $n_0(\bs r,t)$. 

So, at least formally, there seems to be a way to establish a valid td FPI scheme based entirely on the RG1-type mappings, featuring td KS equations to be solved iteratively in a given time interval.
However, where is the catch? The problem is that this FPI scheme is an \textbf{\textit{ad hoc} construct}, not backed by any 
variational or stationarity principle. This means that there is no reason to expect that the FPI procedure converges, at least in principle. Thus, the inevitable conclusion is that also the RG1-type mapping based derivation of td KS equations
does not constitute a valid foundation of TDDFT.
Let us note that an analogous mapping based foundation could also be conceived for the 
(static) KS equations of ordinary DFT, as is briefly discussed in App.~A.

Remarkably, the FPI concept has hardly been communicated in the TDDFT literature and therefore not been exposed to critical examinations. A notable exception, though, is a brief discussion in the textbook by Ullrich~\cite{Ullrich:2012}. In fact, the FPI scheme was not considered to be a practical computational tool, 
let alone used in actual applications.    
A handy workaround seemed available, namely, on the fly' (OTF) time propagation of the td KS equations~(\ref{eq:tdksfp}),
being not only a practical computational procedure but also dispensing with the inconvenient convergence issue of the 
FPI mode. So let us inspect the OTF procedure, where the suppressed FPI problems reappear in a different form.


On the first glance, Eqs.~(\ref{eq:tdksfp}) look like a set of explicit first-order 
differential equations amenable to OTF time propagation. 
However, as our analysis of the td rKS scheme has shown, the determination of $v_{KS}[n](\bs r,t)$ (and by extension
also of $v_{xc}[n](\bs r,t)$) at a time $t$ requires besides $n(\bs r,t)$ also the first and second time derivatives,
$\dot n(\bs r,t)$ and $\ddot n(\bs r,t)$. This means that, strictly speaking, the td KS equations~(\ref{eq:tdksfp})
represent a set of implicit second-order differential equations that cannot be propagated in a first-order OTF fashion.
Of course, in the adiabatic approximation, where the xc PF depends on time only via the time-dependence of the density, $n(\bs r,t)$, an OTF propagation of the td KS equations is feasible. But here the question is: an approximation to what exactly, as there is no underlying exact approach? So, also the OTF procedure does not solve the problem inherent to the FPI concept.

Let us note that in the FPI mode, starting the iterations with a trial density trajectory, $\tilde n_0(\bs r,t),\, t_0 \leq t \leq t_1$, there is of course no problem in using common propagation techniques to solve the td KS equations at a given iteration step. At each step the entire density trajectory of the preceding step is available and can be used as input to determine the xc PF required in the current step for any time within the given interval.
As already stated, the actual problem of the FPI procedure is the lack of a convergence rationale.
Thus, neither the original FPI scheme nor the OTF shortcut salvage the lost cause of the mapping-based foundation of 
td KS equations.

\section{Summary and concluding remarks}

The original foundation of TDDFT by Runge and Gross (RG) from 1984 has proven to be invalid.
Here a central pillar was a stationarity principle for an action-integral functional (AIF)
established via the RG1 mapping theorem. According to our analysis, this AIF  must be seen as an irreparable misconstruction, a finding which, however, has not fully been recognized so far.

That there were problems with the RG foundation was realized almost 30 years ago, and it was subsequently abandoned, though without a conclusive analysis of the underlying problems.
What is offered as a substitute - though so far not authoritatively communicated via a second rigorous founding paper - is a direct route to td KS equations without involving a stationarity principle. Based on
two RG1-type mapping theorems, namely the original RG1 mapping for interacting electron systems and
a corresponding mapping for non-interacting particles, there is indeed a formally correct way to postulate 
td KS equations in the form of a fixed-point iteration (FPI) scheme supposed to converge to the density trajectory of
the system under consideration. But there is a catch: The FPI scheme is merely an \textit{ad hoc} construct not backed by
a variational or stationarity principle as a guarantor of the convergence of the FPI procedure. So also the 
mapping/FPI based td KS concept cannot be seen as a valid reconstitution of the TDDFT foundations - though this view does not yet seem to have reached the TDDFT community.
 
In fact, the FPI mode of dealing with the td KS equations was hardly ever discussed, let alone used in practice.   
On the fly' (OTF) propagation seemed to be a legitimate and viable computational method, generating directly the
desired fixed-point density trajectory. However, the problem at the core of the FPI scheme reappears here in another form.
While OTF propagation is possible if the xc potential functional depends only on the td density function, as  
is the case in the adiabatic approximation, one must realize that the putative exact xc potential functional depends also on the first and second time derivative of the density function, which means that the td KS equations constitute a
system of implicit second-order differential equations that cannot be solved by first-order type time propagation. 
Obviously, OTF propagation does not offer a way out of the FPI mode deficiency - again a view not yet taken into serious consideration by the TDDFT architects.    

In view of the critical review given here one has to face the conclusion that there is currently no valid justification for TDDFT, and any  
expectations of finding a remedy here are unfounded. Rather, one should abandon the idea of a formally exact method for dealing with the time-evolution of a many-electron quantum system on the level of electron density functions, and deriving thereof an, in principle, exact way of extending the ground-state DFT approach to electron excitations.   
This means that the LR-TDDFT computational schemes must be seen as what they were in their beginning some 40 years ago:  
pragmatical modifications of the RPA obtained by referring to the KS equations of static DFT augmented by a time-dependent external potential rather than the corresponding  
HF equations. While this may result in empirically improved results for singly excited states, the physical significance 
of such improvements is questionable.

\section*{Acknowledgements}
\noindent
Thanks are due to Krishna Nandipati for the incentive and invitation to give a lecture 
on the status of the TDDFT foundations at the Joint Physical Chemistry Seminar at the Heidelberg University in May 2023.
A critical reading of the manuscript and helpful comments by Marco Bauer, Lorenz S. Cederbaum, and Andreas Dreuw are gratefully acknowledged.
\appendix
\renewcommand{\theequation}{A.\arabic{equation}}
\setcounter{equation}{0}
\section*{Appendix A: Mapping based derivation of the ordinary KS equations}

The concept of a mapping-based foundation of the time-dependent KS equations addressed in Sec.~IV
can be contrasted with an essentially analogous procedure in the more transparent case of static DFT.
This shall be discussed and analysed in the following.

As discussed in Sec.~III.B, the (spin-free) KS equations~(\ref{eq:kseq}),
\begin{equation}
\label{eq:kseeq}
\{-\half \nabla^2 + u(\bs{r}) + J[n](\bs{r}) 
                   + v_{xc}[n](\bs{r})\}\, \phi_k(\bs{r}) = \epsilon_k \phi_k(\bs{r}),\,\, k=1,\dots,N/2
\end{equation}
are obtained by applying the variational principle for the energy density functional of the considered many-electron system 
at the orbital level underlying the electron density. 
Here, $u(\bs r)$ is the one-particle potential of the considered system, and  
\begin{equation}
\label{eq:def1vxc}
 v_{xc}[n](\bs{r}) = \frac{\delta E_{xc}[n]}{\delta n(\bs{r})}
\end{equation} 
is the xc potential functional (PF) deriving from the xc energy functional~(\ref{eq:ksxcfun}).
Together with the relation
\begin{equation}
\label{eq:orb2dens}
n(\bs r) = 2 \sum_{k}^{N/2} |\phi_k(\bs r)|^2
\end{equation}
the KS equations establish an iterative computational scheme, where the ground-state density $n_0(\bs r)$
of the considered system is the desired solution ('fixed point').

Now let us suppose that a HKII variational principle is not available, and one 
has to resort, firstly, to the HKI mapping theorem~(\ref{eq:n2v}) for interacting many-electron systems,
\begin{equation}
n(\bs{r}) \longrightarrow v_{ext}[n](\bs{r}) + c
\end{equation}
(adapting here the notation of the PF to the TDDFT usage),
and, secondly, an analogous mapping for non-interacting particles,
\begin{equation}
n(\bs{r}) \longrightarrow v_{KS}[n](\bs{r}) + c
\end{equation}
Using the KS PF, $v_{KS}[n]$, for a given density, $n(\bs r)$, in the KS-type orbital equations  
\begin{equation}
\{-\half \nabla^2 + v_{KS}[n](\bs{r})\}\, \phi_k(\bs{r}) = \epsilon_k \phi_k(\bs{r}),\,\, k=1,\dots,N/2
\end{equation}
reproduces the density according to $n(\bs r) = 2 \sum_{k}^{N/2} |\phi_k(\bs r)|^2$.

It should be noted that the non-interaction concept could be made even more stringent
by limiting oneself to a single orbital equation,
\begin{equation}
\{-\half \nabla^2 + v_{KS}[n](\bs{r})\}\, \phi(\bs{r}) = \epsilon \phi(\bs{r})
\end{equation}
as is discussed in Sec.~III.C under the designation rKS (radical KS).  
Here, the corresponding PF can explicitly be constructed, yielding the simple expression~\cite{sch07:022513}
\begin{equation}
v_{KS}[n](\bs{r}) = \frac{\nabla^2 \sqrt{n(\bs{r})}}{2 \sqrt{n(\bs{r})}} + c
\end{equation}
At this rKS level, the triviality of the KS PF and the non-interaction mapping is manifest.

Equipped with these two mapping theorems and the associated PFs 
one can define a non-trivial xc PF according to
\begin{equation}
\label{eq:def2vxc}
\tilde v_{xc}[n](\bs r) \equiv v_{KS}[n](\bs r) - v_{ext}[n](\bs r) - J[n](\bs r)
\end{equation}
and postulate KS equations for the system under consideration,
\begin{equation}
\label{eq:ksexq}
\{-\half \nabla^2 + u(\bs{r}) + J[n](\bs{r}) 
                   + \tilde v_{xc}[n](\bs{r})\}\, \phi_k(\bs{r}) = \epsilon_k \phi_k(\bs{r}),\,\, k=1,\dots,N/2
\end{equation}
Together with Eq.~(\ref{eq:orb2dens}) these equations constitute a fixed-point iteration (FPI) scheme  
supposed to yield the desired density $n_0(\bs r)$ of the considered system as the fixed point.
To understand the rationale here, Eqs.~(\ref{eq:ksexq}) can be written more explicitly as
\begin{equation}
\{-\half \nabla^2 + u(\bs{r}) + v_{KS}[n](\bs{r}) 
                   - v_{ext}[n](\bs{r})\}\, \phi_k(\bs{r}) = \epsilon_k \phi_k(\bs{r}),\,\, k=1,\dots,N/2
\end{equation}
Consequently, when $n(\bs r)$ approaches $n_0(\bs r)$, the external potential cancels $u(\bs{r})$, since $v_{ext}[n_0](\bs{r}) = u(\bs{r}) + c$, and the remaining potential $v_{KS}[n_0](\bs{r})$ reproduces $n_0(\bs r)$. 

However, two questions remain. The first concerns the convergence of that FPI procedure, the second is about the  
relation between the 'canonical' and the mapping-based xc PF, $v_{xc}[n](\bs r)$ and $\tilde v_{xc}[n](\bs r)$, 
respectively.

Let us first consider the second question. Here it is important to note that
any density, say $\tilde n(\bs r)$, can be made the FP density in either variant of the KS equations 
by just replacing the original potential $u(\bs r)$ with another specific potential, $v_{ext}[\tilde n](\bs r)$, in Eqs. (\ref{eq:kseeq}) 
and Eqs.~(\ref{eq:ksexq}), respectively.
The fact that both KS variants have the same FP density for arbitrary densities $\tilde n(\bs r)$
suggests the conclusion that they are in fact identical, and that, in particular, 
\begin{equation}
v_{xc}[n](\bs{r}) = \tilde v_{xc}[n](\bs{r}) = v_{KS}[n](\bs{r}) - v_{ext}[n](\bs{r}) - J[n](\bs{r})
\end{equation}
This is an interesting finding, as it also establishes via Eq.~(\ref{eq:def1vxc}) the connection to the
(HKI based) energy functional $E[n]$, and, specifically, the energy, $E[n_0]$, associated with FP solution $n_0(\bs r)$
of the original system. But it should be recalled that the canonical KS equations are based on the HKII variational principle,
which here also furnishes the in-principle convergence of the FPI procedure. 
On their own, without the HKII or a related variational principle in the background, the mapping-based KS equations
are unfounded, as there is no reason to expect the FPI scheme to converge.


\begin{thebibliography}{1}
\expandafter\ifx\csname natexlab\endcsname\relax\def\natexlab#1{#1}\fi
\expandafter\ifx\csname bibnamefont\endcsname\relax
  \def\bibnamefont#1{#1}\fi
\expandafter\ifx\csname bibfnamefont\endcsname\relax
  \def\bibfnamefont#1{#1}\fi
\expandafter\ifx\csname citenamefont\endcsname\relax
  \def\citenamefont#1{#1}\fi
\expandafter\ifx\csname url\endcsname\relax
  \def\url#1{\texttt{#1}}\fi
\expandafter\ifx\csname urlprefix\endcsname\relax\def\urlprefix{URL }\fi
\providecommand{\bibinfo}[2]{#2}
\providecommand{\eprint}[2][]{\url{#2}}

\bibitem[{\citenamefont{Burke et~al.}(2005)\citenamefont{Burke, Werschnik, and
  Gross}}]{bur05:062206}
\bibinfo{author}{\bibfnamefont{K.}~\bibnamefont{Burke}},
  \bibinfo{author}{\bibfnamefont{J.}~\bibnamefont{Werschnik}},
  \bibnamefont{and} \bibinfo{author}{\bibfnamefont{E.~K.~U.}
  \bibnamefont{Gross}}, \bibinfo{journal}{J. Chem. Phys.}
  \textbf{\bibinfo{volume}{123}}, \bibinfo{pages}{062206}
  (\bibinfo{year}{2005}).

\bibitem[{\citenamefont{Elliott et~al.}(2008)\citenamefont{Elliott, Burke, and
  Furche}}]{ell08:91}
\bibinfo{author}{\bibfnamefont{P.}~\bibnamefont{Elliott}},
  \bibinfo{author}{\bibfnamefont{K.}~\bibnamefont{Burke}}, \bibnamefont{and}
  \bibinfo{author}{\bibfnamefont{F.}~\bibnamefont{Furche}},
  \bibinfo{journal}{Rev. Comp. Chem.} \textbf{\bibinfo{volume}{26}},
  \bibinfo{pages}{91} (\bibinfo{year}{2008}).

\bibitem[{\citenamefont{Casida}(2009)}]{cas09:3}
\bibinfo{author}{\bibfnamefont{M.~E.} \bibnamefont{Casida}},
  \bibinfo{journal}{J. Mol. Struct.: THEOCHEM} \textbf{\bibinfo{volume}{914}},
  \bibinfo{pages}{3} (\bibinfo{year}{2009}).

\bibitem[{\citenamefont{Gross and Maitra}(2012)}]{gro12:53}
\bibinfo{author}{\bibfnamefont{E.~K.~U.} \bibnamefont{Gross}} \bibnamefont{and}
  \bibinfo{author}{\bibfnamefont{N.~T.} \bibnamefont{Maitra}}, in
  \emph{\bibinfo{booktitle}{Fundamentals of Time-dependent Density Functional
  Theory}}, edited by \bibinfo{editor}{\bibfnamefont{M.~A.~L.}
  \bibnamefont{Marques}}, \bibinfo{editor}{\bibfnamefont{N.~T.}
  \bibnamefont{Maitra}}, \bibinfo{editor}{\bibfnamefont{F.~M.~S.}
  \bibnamefont{Nogueira}}, \bibnamefont{and}
  \bibinfo{editor}{\bibfnamefont{E.~K.~U.} \bibnamefont{Gross}}
  (\bibinfo{publisher}{Springer}, \bibinfo{address}{Heidelberg},
  \bibinfo{year}{2012}).

\bibitem[{\citenamefont{Ruggenthaler et~al.}(2015)\citenamefont{Ruggenthaler,
  Penz, and van Leeuwen}}]{rug15:203202}
\bibinfo{author}{\bibfnamefont{M.}~\bibnamefont{Ruggenthaler}},
  \bibinfo{author}{\bibfnamefont{M.}~\bibnamefont{Penz}}, \bibnamefont{and}
  \bibinfo{author}{\bibfnamefont{R.}~\bibnamefont{van Leeuwen}},
  \bibinfo{journal}{J. Phys.: Condens. Matter} \textbf{\bibinfo{volume}{27}},
  \bibinfo{pages}{203202} (\bibinfo{year}{2015}).

\bibitem[{\citenamefont{Maitra}(2016)}]{mai16:220901}
\bibinfo{author}{\bibfnamefont{N.~T.} \bibnamefont{Maitra}},
  \bibinfo{journal}{J. Chem. Phys.} \textbf{\bibinfo{volume}{144}},
  \bibinfo{pages}{220901} (\bibinfo{year}{2016}).

\bibitem[{\citenamefont{Huix-Rotllant et~al.}(2021)\citenamefont{Huix-Rotllant,
  Ferr\'{e}, and Barbatti}}]{hui21:15}
\bibinfo{author}{\bibfnamefont{M.}~\bibnamefont{Huix-Rotllant}},
  \bibinfo{author}{\bibfnamefont{N.}~\bibnamefont{Ferr\'{e}}},
  \bibnamefont{and} \bibinfo{author}{\bibfnamefont{M.}~\bibnamefont{Barbatti}},
  in \emph{\bibinfo{booktitle}{Quantum Chemistry and Dynamics of Excited
  States: Methods and Applications}}, edited by
  \bibinfo{editor}{\bibfnamefont{L.}~\bibnamefont{Gonz\'{a}les}}
  \bibnamefont{and} \bibinfo{editor}{\bibfnamefont{R.}~\bibnamefont{Lindh}}
  (\bibinfo{publisher}{Wiley}, \bibinfo{address}{Hoboken, New Jersey},
  \bibinfo{year}{2021}).

\bibitem[{\citenamefont{Engel and Dreizler}(2011)}]{Engel:2011}
\bibinfo{author}{\bibfnamefont{E.}~\bibnamefont{Engel}} \bibnamefont{and}
  \bibinfo{author}{\bibfnamefont{R.}~\bibnamefont{Dreizler}},
  \emph{\bibinfo{title}{Density Functional Theory}}
  (\bibinfo{publisher}{Springer}, \bibinfo{address}{Berlin Heidelberg},
  \bibinfo{year}{2011}).

\bibitem[{\citenamefont{Ullrich}(2012)}]{Ullrich:2012}
\bibinfo{author}{\bibfnamefont{C.~A.} \bibnamefont{Ullrich}},
  \emph{\bibinfo{title}{Time-Dependent Density-Functional Theory}}
  (\bibinfo{publisher}{Oxford University Press}, \bibinfo{address}{Oxford},
  \bibinfo{year}{2012}).

\bibitem[{\citenamefont{Silva-Junior et~al.}(2008)\citenamefont{Silva-Junior,
  Schreiber, Sauer, and Thiel}}]{sil08:104103}
\bibinfo{author}{\bibfnamefont{M.~R.} \bibnamefont{Silva-Junior}},
  \bibinfo{author}{\bibfnamefont{M.}~\bibnamefont{Schreiber}},
  \bibinfo{author}{\bibfnamefont{S.~P.~A.} \bibnamefont{Sauer}},
  \bibnamefont{and} \bibinfo{author}{\bibfnamefont{W.}~\bibnamefont{Thiel}},
  \bibinfo{journal}{J. Chem. Phys.} \textbf{\bibinfo{volume}{129}},
  \bibinfo{pages}{104103} (\bibinfo{year}{2008}).

\bibitem[{\citenamefont{Jacquemin et~al.}(2009)\citenamefont{Jacquemin,
  Wathelet, Perp\'{e}te, and Adamo}}]{jac09:2420}
\bibinfo{author}{\bibfnamefont{D.}~\bibnamefont{Jacquemin}},
  \bibinfo{author}{\bibfnamefont{V.}~\bibnamefont{Wathelet}},
  \bibinfo{author}{\bibfnamefont{E.~A.} \bibnamefont{Perp\'{e}te}},
  \bibnamefont{and} \bibinfo{author}{\bibfnamefont{C.}~\bibnamefont{Adamo}},
  \bibinfo{journal}{J.~Chem.~Theory~Comput.} \textbf{\bibinfo{volume}{5}},
  \bibinfo{pages}{2420} (\bibinfo{year}{2009}).

\bibitem[{\citenamefont{Caricato et~al.}(2010)\citenamefont{Caricato, Trucks,
  Frisch, and Wiberg}}]{car10:370}
\bibinfo{author}{\bibfnamefont{M.}~\bibnamefont{Caricato}},
  \bibinfo{author}{\bibfnamefont{G.~W.} \bibnamefont{Trucks}},
  \bibinfo{author}{\bibfnamefont{M.~J.} \bibnamefont{Frisch}},
  \bibnamefont{and} \bibinfo{author}{\bibfnamefont{K.~B.}
  \bibnamefont{Wiberg}}, \bibinfo{journal}{J.~Chem.~Theory~Comput.}
  \textbf{\bibinfo{volume}{6}}, \bibinfo{pages}{370} (\bibinfo{year}{2010}).

\bibitem[{\citenamefont{Hohenberg and Kohn}(1964)}]{hoh64:864}
\bibinfo{author}{\bibfnamefont{P.}~\bibnamefont{Hohenberg}} \bibnamefont{and}
  \bibinfo{author}{\bibfnamefont{W.}~\bibnamefont{Kohn}},
  \bibinfo{journal}{Phys. Rev.} \textbf{\bibinfo{volume}{136}},
  \bibinfo{pages}{B 864} (\bibinfo{year}{1964}).

\bibitem[{\citenamefont{Kohn and Sham}(1965)}]{koh65:1133}
\bibinfo{author}{\bibfnamefont{W.}~\bibnamefont{Kohn}} \bibnamefont{and}
  \bibinfo{author}{\bibfnamefont{L.~J.} \bibnamefont{Sham}},
  \bibinfo{journal}{Phys. Rev.} \textbf{\bibinfo{volume}{140}},
  \bibinfo{pages}{A 1133} (\bibinfo{year}{1965}).

\bibitem[{\citenamefont{Zangwil and Soven}(1980)}]{zan80:1561}
\bibinfo{author}{\bibfnamefont{A.}~\bibnamefont{Zangwil}} \bibnamefont{and}
  \bibinfo{author}{\bibfnamefont{P.}~\bibnamefont{Soven}},
  \bibinfo{journal}{Phys. Rev. A} \textbf{\bibinfo{volume}{21}},
  \bibinfo{pages}{1561} (\bibinfo{year}{1980}).

\bibitem[{\citenamefont{Runge and Gross}(1984)}]{run84:997}
\bibinfo{author}{\bibfnamefont{E.}~\bibnamefont{Runge}} \bibnamefont{and}
  \bibinfo{author}{\bibfnamefont{E.~K.~U.} \bibnamefont{Gross}},
  \bibinfo{journal}{Phys. Rev. Lett.} \textbf{\bibinfo{volume}{52}},
  \bibinfo{pages}{997} (\bibinfo{year}{1984}).

\bibitem[{\citenamefont{Schirmer and Dreuw}(2007)}]{sch07:022513}
\bibinfo{author}{\bibfnamefont{J.}~\bibnamefont{Schirmer}} \bibnamefont{and}
  \bibinfo{author}{\bibfnamefont{A.}~\bibnamefont{Dreuw}},
  \bibinfo{journal}{Phys. Rev. A} \textbf{\bibinfo{volume}{75}},
  \bibinfo{pages}{022513} (\bibinfo{year}{2007}).

\bibitem[{\citenamefont{Maitra et~al.}(2008)\citenamefont{Maitra, van Leeuwen,
  and Burke}}]{mai08:056501}
\bibinfo{author}{\bibfnamefont{N.~T.} \bibnamefont{Maitra}},
  \bibinfo{author}{\bibfnamefont{R.}~\bibnamefont{van Leeuwen}},
  \bibnamefont{and} \bibinfo{author}{\bibfnamefont{K.}~\bibnamefont{Burke}},
  \bibinfo{journal}{Phys.~Rev.~A} \textbf{\bibinfo{volume}{78}},
  \bibinfo{pages}{056501} (\bibinfo{year}{2008}).

\bibitem[{\citenamefont{Schirmer and Dreuw}(2008)}]{sch08:056502}
\bibinfo{author}{\bibfnamefont{J.}~\bibnamefont{Schirmer}} \bibnamefont{and}
  \bibinfo{author}{\bibfnamefont{A.}~\bibnamefont{Dreuw}},
  \bibinfo{journal}{Phys. Rev. A} \textbf{\bibinfo{volume}{78}},
  \bibinfo{pages}{056502} (\bibinfo{year}{2008}).

\bibitem[{\citenamefont{Schirmer}(2012)}]{sch12:012514}
\bibinfo{author}{\bibfnamefont{J.}~\bibnamefont{Schirmer}},
  \bibinfo{journal}{Phys.~Rev. A} \textbf{\bibinfo{volume}{86}},
  \bibinfo{pages}{012514} (\bibinfo{year}{2012}).

\bibitem[{\citenamefont{Schirmer et~al.}(2001)\citenamefont{Schirmer,
  Braunstein, Lee, and McKoy}}]{sch95:105}
\bibinfo{author}{\bibfnamefont{J.}~\bibnamefont{Schirmer}},
  \bibinfo{author}{\bibfnamefont{M.}~\bibnamefont{Braunstein}},
  \bibinfo{author}{\bibfnamefont{M.-T.} \bibnamefont{Lee}}, \bibnamefont{and}
  \bibinfo{author}{\bibfnamefont{V.}~\bibnamefont{McKoy}}, in
  \emph{\bibinfo{booktitle}{VUV and Soft X-Ray Photoionization}}, edited by
  \bibinfo{editor}{\bibfnamefont{U.}~\bibnamefont{Becker}} \bibnamefont{and}
  \bibinfo{editor}{\bibfnamefont{D.~A.} \bibnamefont{Shirley}}
  (\bibinfo{publisher}{Plenum Press}, \bibinfo{address}{New York},
  \bibinfo{year}{2001}), p. \bibinfo{pages}{105}.

\bibitem[{\citenamefont{Schirmer}(2018)}]{Schirm:2018}
\bibinfo{author}{\bibfnamefont{J.}~\bibnamefont{Schirmer}},
  \emph{\bibinfo{title}{Many-Body Methods for Atoms, Molecules and Clusters}}
  (\bibinfo{publisher}{Springer}, \bibinfo{address}{Heidelberg},
  \bibinfo{year}{2018}).

\bibitem[{\citenamefont{McLachlan and Ball}(1964)}]{mcl64:844}
\bibinfo{author}{\bibfnamefont{A.~D.} \bibnamefont{McLachlan}}
  \bibnamefont{and} \bibinfo{author}{\bibfnamefont{M.~A.} \bibnamefont{Ball}},
  \bibinfo{journal}{Rev. Mod. Phys.} \textbf{\bibinfo{volume}{36}},
  \bibinfo{pages}{844} (\bibinfo{year}{1964}).

\bibitem[{\citenamefont{Dunning and McKoy}(1967)}]{dun67:1735}
\bibinfo{author}{\bibfnamefont{T.~H.} \bibnamefont{Dunning}} \bibnamefont{and}
  \bibinfo{author}{\bibfnamefont{V.}~\bibnamefont{McKoy}}, \bibinfo{journal}{J.
  Chem. Phys.} \textbf{\bibinfo{volume}{47}}, \bibinfo{pages}{1735}
  (\bibinfo{year}{1967}).

\bibitem[{\citenamefont{Rowe}(1968)}]{row68:153}
\bibinfo{author}{\bibfnamefont{D.~J.} \bibnamefont{Rowe}},
  \bibinfo{journal}{Rev. Mod. Phys.} \textbf{\bibinfo{volume}{40}},
  \bibinfo{pages}{153} (\bibinfo{year}{1968}).

\bibitem[{\citenamefont{Ring and Schuck}(2000)}]{Ring:1980}
\bibinfo{author}{\bibfnamefont{P.}~\bibnamefont{Ring}} \bibnamefont{and}
  \bibinfo{author}{\bibfnamefont{P.}~\bibnamefont{Schuck}},
  \emph{\bibinfo{title}{The Nuclear Many-Body Problem}}
  (\bibinfo{publisher}{Springer}, \bibinfo{address}{Heidelberg},
  \bibinfo{year}{2000}).

\bibitem[{\citenamefont{Casida}(1995)}]{cas95:155}
\bibinfo{author}{\bibfnamefont{M.~E.} \bibnamefont{Casida}}, in
  \emph{\bibinfo{booktitle}{Recent Advances in Density Functional Theory, Part
  I}}, edited by \bibinfo{editor}{\bibfnamefont{D.~P.} \bibnamefont{Chong}}
  (\bibinfo{publisher}{World Scientific}, \bibinfo{address}{Singapore},
  \bibinfo{year}{1995}).

\bibitem[{\citenamefont{Bauernschmitt and Ahlrichs}(1996)}]{bau96:454}
\bibinfo{author}{\bibfnamefont{R.}~\bibnamefont{Bauernschmitt}}
  \bibnamefont{and} \bibinfo{author}{\bibfnamefont{R.}~\bibnamefont{Ahlrichs}},
  \bibinfo{journal}{Chem. Phys. Lett.} \textbf{\bibinfo{volume}{256}},
  \bibinfo{pages}{454} (\bibinfo{year}{1996}).

\bibitem[{\citenamefont{Casida et~al.}(1998)\citenamefont{Casida, Jamorski,
  Casida, and Salahub}}]{cas98:4439}
\bibinfo{author}{\bibfnamefont{M.~E.} \bibnamefont{Casida}},
  \bibinfo{author}{\bibfnamefont{C.}~\bibnamefont{Jamorski}},
  \bibinfo{author}{\bibfnamefont{K.~C.} \bibnamefont{Casida}},
  \bibnamefont{and} \bibinfo{author}{\bibfnamefont{D.~R.}
  \bibnamefont{Salahub}}, \bibinfo{journal}{J. Chem. Phys.}
  \textbf{\bibinfo{volume}{108}}, \bibinfo{pages}{4439} (\bibinfo{year}{1998}).

\bibitem[{\citenamefont{Tozer and Handy}(1998)}]{toz98:10180}
\bibinfo{author}{\bibfnamefont{D.~J.} \bibnamefont{Tozer}} \bibnamefont{and}
  \bibinfo{author}{\bibfnamefont{N.~C.} \bibnamefont{Handy}},
  \bibinfo{journal}{J. Chem. Phys.} \textbf{\bibinfo{volume}{109}},
  \bibinfo{pages}{10180} (\bibinfo{year}{1998}).

\bibitem[{\citenamefont{Tozer et~al.}(1999)\citenamefont{Tozer, Amos, Handy,
  Roos, and Serrano-Andr{\'e}s}}]{toz99:859}
\bibinfo{author}{\bibfnamefont{D.~J.} \bibnamefont{Tozer}},
  \bibinfo{author}{\bibfnamefont{R.~D.} \bibnamefont{Amos}},
  \bibinfo{author}{\bibfnamefont{N.~C.} \bibnamefont{Handy}},
  \bibinfo{author}{\bibfnamefont{B.}~\bibnamefont{Roos}}, \bibnamefont{and}
  \bibinfo{author}{\bibfnamefont{L.}~\bibnamefont{Serrano-Andr{\'e}s}},
  \bibinfo{journal}{Mol. Phys.} \textbf{\bibinfo{volume}{97}},
  \bibinfo{pages}{859} (\bibinfo{year}{1999}).

\bibitem[{\citenamefont{Dreuw et~al.}(2003)\citenamefont{Dreuw, Weisman, and
  Head-Gordon}}]{dre03:2943}
\bibinfo{author}{\bibfnamefont{A.}~\bibnamefont{Dreuw}},
  \bibinfo{author}{\bibfnamefont{J.~L.} \bibnamefont{Weisman}},
  \bibnamefont{and}
  \bibinfo{author}{\bibfnamefont{M.}~\bibnamefont{Head-Gordon}},
  \bibinfo{journal}{J. Chem. Phys.} \textbf{\bibinfo{volume}{119}},
  \bibinfo{pages}{2943} (\bibinfo{year}{2003}).

\bibitem[{\citenamefont{Dreuw and Head-Gordon}(2004)}]{dre04:4007}
\bibinfo{author}{\bibfnamefont{A.}~\bibnamefont{Dreuw}} \bibnamefont{and}
  \bibinfo{author}{\bibfnamefont{M.}~\bibnamefont{Head-Gordon}},
  \bibinfo{journal}{J. Am. Chem. Soc.} \textbf{\bibinfo{volume}{126}},
  \bibinfo{pages}{4007} (\bibinfo{year}{2004}).

\bibitem[{\citenamefont{Sobolewski and Domcke}(2003)}]{sob03:73}
\bibinfo{author}{\bibfnamefont{A.~L.} \bibnamefont{Sobolewski}}
  \bibnamefont{and} \bibinfo{author}{\bibfnamefont{W.}~\bibnamefont{Domcke}},
  \bibinfo{journal}{Chem.~Phys} \textbf{\bibinfo{volume}{294}},
  \bibinfo{pages}{73} (\bibinfo{year}{2003}).

\bibitem[{\citenamefont{Parr and Yang}(1989)}]{Parr:1989}
\bibinfo{author}{\bibfnamefont{R.~G.} \bibnamefont{Parr}} \bibnamefont{and}
  \bibinfo{author}{\bibfnamefont{W.}~\bibnamefont{Yang}},
  \emph{\bibinfo{title}{Density-Functional Theory of Atoms and Molecules}}
  (\bibinfo{publisher}{Oxford University Press - Clarendon Press},
  \bibinfo{address}{New York, Oxford}, \bibinfo{year}{1989}).

\bibitem[{\citenamefont{Nagy}(1998)}]{nag98:1}
\bibinfo{author}{\bibfnamefont{A.}~\bibnamefont{Nagy}},
  \bibinfo{journal}{Phys.Rep.} \textbf{\bibinfo{volume}{298}},
  \bibinfo{pages}{1} (\bibinfo{year}{1998}).

\bibitem[{\citenamefont{Levy et~al.}(1984)\citenamefont{Levy, Perdew, and
  Sahni}}]{lev84:2745}
\bibinfo{author}{\bibfnamefont{M.}~\bibnamefont{Levy}},
  \bibinfo{author}{\bibfnamefont{J.~P.} \bibnamefont{Perdew}},
  \bibnamefont{and} \bibinfo{author}{\bibfnamefont{V.}~\bibnamefont{Sahni}},
  \bibinfo{journal}{Phys. Rev. A} \textbf{\bibinfo{volume}{30}},
  \bibinfo{pages}{2745} (\bibinfo{year}{1984}).

\bibitem[{\citenamefont{Holas and March}(1991)}]{hol91:5521}
\bibinfo{author}{\bibfnamefont{A.}~\bibnamefont{Holas}} \bibnamefont{and}
  \bibinfo{author}{\bibfnamefont{N.~H.} \bibnamefont{March}},
  \bibinfo{journal}{Phys. Rev. A} \textbf{\bibinfo{volume}{44}},
  \bibinfo{pages}{5521} (\bibinfo{year}{1991}).

\bibitem[{\citenamefont{Gross et~al.}(1996)\citenamefont{Gross, Dobson, and
  Petersilka}}]{Nalewajski:1996}
\bibinfo{author}{\bibfnamefont{E.~K.~U.} \bibnamefont{Gross}},
  \bibinfo{author}{\bibfnamefont{J.~F.} \bibnamefont{Dobson}},
  \bibnamefont{and}
  \bibinfo{author}{\bibfnamefont{M.}~\bibnamefont{Petersilka}}, in
  \emph{\bibinfo{booktitle}{Density Functional Theory}}, edited by
  \bibinfo{editor}{\bibfnamefont{R.~F.} \bibnamefont{Nalewajski}}
  (\bibinfo{publisher}{Springer}, \bibinfo{address}{New York},
  \bibinfo{year}{1996}).

\bibitem[{\citenamefont{Rajagopal}(1996)}]{raj96:3916}
\bibinfo{author}{\bibfnamefont{A.~K.} \bibnamefont{Rajagopal}},
  \bibinfo{journal}{Phys. Rev. A} \textbf{\bibinfo{volume}{54}},
  \bibinfo{pages}{3916} (\bibinfo{year}{1996}).

\bibitem[{\citenamefont{van Leeuwen}(1998)}]{lee98:1280}
\bibinfo{author}{\bibfnamefont{R.}~\bibnamefont{van Leeuwen}},
  \bibinfo{journal}{Phys. Rev. Lett.} \textbf{\bibinfo{volume}{80}},
  \bibinfo{pages}{1280} (\bibinfo{year}{1998}).

\bibitem[{\citenamefont{Harbola and Banerjee}(1999)}]{har99:5101}
\bibinfo{author}{\bibfnamefont{M.~K.} \bibnamefont{Harbola}} \bibnamefont{and}
  \bibinfo{author}{\bibfnamefont{A.}~\bibnamefont{Banerjee}},
  \bibinfo{journal}{Phys. Rev. A} \textbf{\bibinfo{volume}{60}},
  \bibinfo{pages}{5101} (\bibinfo{year}{1999}).

\bibitem[{\citenamefont{van Leeuwen}(2001)}]{lee01:1969}
\bibinfo{author}{\bibfnamefont{R.}~\bibnamefont{van Leeuwen}},
  \bibinfo{journal}{Int. J. Mod. Phys. B} \textbf{\bibinfo{volume}{14}},
  \bibinfo{pages}{1969} (\bibinfo{year}{2001}).

\bibitem[{\citenamefont{Vignale}(2008)}]{vig08:062511}
\bibinfo{author}{\bibfnamefont{G.}~\bibnamefont{Vignale}},
  \bibinfo{journal}{Phys. Rev. A} \textbf{\bibinfo{volume}{77}},
  \bibinfo{pages}{062511} (\bibinfo{year}{2008}).

\bibitem[{\citenamefont{van Leeuwen}(1999)}]{lee99:3863}
\bibinfo{author}{\bibfnamefont{R.}~\bibnamefont{van Leeuwen}},
  \bibinfo{journal}{Phys. Rev. Lett.} \textbf{\bibinfo{volume}{82}},
  \bibinfo{pages}{3863} (\bibinfo{year}{1999}).

\bibitem[{\citenamefont{Marques and Gross}(2004)}]{mar04:427}
\bibinfo{author}{\bibfnamefont{M.~A.~L.} \bibnamefont{Marques}}
  \bibnamefont{and} \bibinfo{author}{\bibfnamefont{E.~K.~U.}
  \bibnamefont{Gross}}, \bibinfo{journal}{Annu. Rev. Phys. Chem.}
  \textbf{\bibinfo{volume}{55}}, \bibinfo{pages}{427} (\bibinfo{year}{2004}).

\end{thebibliography}
\end{document}